\documentclass[11pt]{article}

\usepackage{authblk,booktabs,setspace,
                crop,inputenc,caption,subcaption,
            array,amscd,amsmath,amssymb,latexsym,
                float,epsfig,wrapfig,graphicx,stfloats,
                multicol,rotating,multirow,dcolumn,
                boxedminipage,makeidx,soul,url,xspace,times}

\usepackage[margin=1in]{geometry} 
\usepackage[all]{xy}
\usepackage[nospace,compress]{cite}
\usepackage{siunitx}

\graphicspath{{FIGS/}}

\newcommand{\comment}[1]{}

\usepackage{xcolor}
\newcommand{\NOTE}[1]{\textcolor{red}{[#1]}}

\usepackage{fancyvrb}

\usepackage{bbold}
\newcommand{\indi}[1]{\mathbb{1}(#1)}

\newcommand{\ie}{i.e.\@\xspace}
\newcommand{\eg}{e.g.\@\xspace}

\def\tableparts#1#2#3{\small{#1 #2 #3}}
\def\colrule{\midrule}
\def\botrule{\bottomrule}

\begin{document}

\title{Single nucleotide polymorphisms that modulate microRNA regulation of gene expression in tumors}

\author[1]{Gary Wilk}
\author[2,3,*]{Rosemary Braun}

\affil[1]{Department of Chemical and Biological Engineering, Northwestern University, Evanston, IL 60208, USA}
\affil[2]{Biostatistics Division, Feinberg School of Medicine, Northwestern University, Chicago, IL 60611, USA}
\affil[3]{Department of Engineering Sciences and Applied Mathematics, Northwestern University, Evanston, IL 60208, USA}
\affil[*]{To whom correspondence should be addressed.  Tel: +312-503-3644; Email: rbraun@northwestern.edu}

\maketitle

\begin{abstract}
Genome-wide association studies (GWAS) have identified single nucleotide polymorphisms (SNPs)
associated with trait diversity and disease susceptibility, yet
the functional properties of many genetic variants and 
their molecular interactions remains unclear.  It has been hypothesized that SNPs in
microRNA binding sites may disrupt gene regulation by microRNAs (miRNAs), short non-coding 
RNAs that bind to mRNA and downregulate the target gene. While a number of studies 
have been conducted to predict the location of SNPs in miRNA binding sites, 
to date there has been no comprehensive analysis of how SNP variants may impact
miRNA regulation of genes.

Here we investigate the functional properties of 
genetic variants and their effects on miRNA regulation of gene expression in cancer. 
Our analysis is motivated by the hypothesis that distinct alleles may cause differential binding (from miRNAs to mRNAs or from transcription factors to DNA) and change the expression of genes. We previously identified pathways---systems of genes conferring specific 
cell functions---that are dysregulated by miRNAs in cancer, by comparing miRNA--pathway 
associations between healthy and tumor tissue. We draw on these results as a starting 
point to assess whether SNPs in genes on dysregulated pathways are responsible for miRNA 
dysregulation of individual genes in tumors. Using an
integrative analysis that incorporates miRNA expression, mRNA expression, and SNP 
genotype data, we identify SNPs that appear to influence the association between miRNAs and
genes, which we term ``regulatory QTLs (regQTLs)'': loci whose alleles impact the
regulation of genes by miRNAs. We describe the method, apply it to analyze four 
cancer types (breast, liver, lung, prostate) using data from The Cancer Genome Atlas (TCGA),
and provide a tool to explore the findings.
\end{abstract}

\section{Introduction}

MicroRNAs (miRNAs) are small noncoding RNA molecules that 
modulate gene expression post-transcriptionally by means of 
complementary base pairing with mRNA transcripts. Through 
recognition of short target motifs (6-8 bases long) on the target mRNA, 
miRNAs bind and down-regulate the expression of the targetted gene. 
Regions flanking the ``seed region'' of the miRNA typically also bind 
the mRNA, creating a stronger annealing between the two RNA molecules. 
This results in the transcript being 
prevented from being translated into protein or degraded in the 
cell~\cite{bartel2004micrornas}.  Because these molecular interactions 
are executed through base pairing, they can be influenced 
by genomic variation; changes in genome sequence may influence 
binding energy and the strength of annealing, or may even abrogate 
miRNA target sites entirely~\cite{chen2008polymorphisms}. 

Polymorphisms constitute approximately 1\% of the human genome 
and contribute to phenotypic diversity and susceptibility to disease. As 
such, large-scale resources to annotate known single nucleotide 
polymorphisms (SNPs) have been constructed, including dbSNP~\cite{sherry2001dbsnp} 
and the International Hapmap Project~\cite{frazer2007second}, 
to describe all known patterns of genetic variation. 
Polymorphisms in miRNA and target site sequences have been
implicated in aberrant miRNA-mRNA interactions and have been
associated with multiple cancers~\cite{salzman2013snping, moszynska2017snps, sethupathy2008microrna},
suggesting a link bewteen genetic variation, miRNA regulation, and
disease.
Typically, 
discoveries of prognostic SNPs come from genome-wide 
association studies (GWAS), which statistically link variants with 
phenotypic traits.
Recent GWAS studies have demonstrated that polymorphisms in 
miRNA binding sites increase the risk of breast~\cite{nicoloso2010single,khan2014microrna}, 
bladder~\cite{yang2008evaluation}, and colon~\cite{naccarati2012polymorphisms,mullany2015snp} 
cancers, among others. In addition, several studies~\cite{salzman2013snping,chen2008polymorphisms} 
have suggested that polymorphisms within miRNA regulatory networks 
affect clinical outcomes and treatment responses.

\enlargethispage{-65.1pt}

In recent years, SNPs and 
their functional effects on miRNA regulation of genes have gained significant 
interest due to observed genetic variation within miRNA networks, and 
several databases and computational tools have been developed 
dedicated toward the study of polymorphic miRNA binding sites. 
These resources include PolymiRTS~\cite{bhattacharya2013polymirts} 
(a database which links polymorphisms with miRNAs and target sites, 
in addition to diseases and biological pathways), 
Patrocles~\cite{hiard2009patrocles} (polymorphisms which are 
predicted to perturb miRNA-gene regulation, including eQTLs and 
Copy Number Variations), and dbSMR~\cite{hariharan2009dbsmr} 
(SNPs around miRNA target sites, genome-wide). These 
resources have improved the search for polymorphic 
binding sites and their potential functional effects in the cell. 
Analogous resources exist to study variation within 
transcription factor (TF) and TF binding sites~\cite{kumar2017snp2tfbs}.


GWAS arrays are not comprehensive, however, and 
often under-sample genomic variants within known miRNA binding regions 
in the genome~\cite{richardson2011genome}.
Additionally, while SNP variants may be predicted to affect
miRNA-gene regulation based on their genomic position, the 
magnitude of the effect is often unclear.
Hence, GWAS
data alone is often insufficient to fully explore the relationship
between genetic variation and miRNA regulation.
Recently, researchers have combined 
GWAS data with separate miRNA expression data in head and neck squamous cell (HNSCC) 
carcinoma to assess variants genome-wide affecting miRNA pathways in 
cancer~\cite{wilkins2017genome}. There, the authors first conducted a GWAS to identify
HSNCC-associated SNP loci, cross-referenced them against putative
miRNA:mRNA binding sites, and confirmed that those miRNAs exhibited
differential expression in the TCGA HSNCC data.  To date, however,
no attempts have been made to directly integrate SNP, miRNA, and gene
expression data from the same samples to identify SNPs that disrupt
miRNA--gene associations, and the functional effects of many
polymorphisms and their molecular interactions remain unknown. 
%
%

To consider the functional effects of SNPs in miRNA networks, several 
criteria are required as outlined in \cite{ryan2010genetic}. These 
criteria include independent association with the phenotype of interest, 
gene expression within the tissue, allelic changes which result in 
differential binding between miRNA and target gene(s), and resultant 
differential target gene expression. Concrete guidelines were 
suggested for future investigations to combine genetic and 
functional evidence for polymorphisms in miRNA target sites 
and human disease~\cite{sethupathy2008microrna}. Follow-up 
functional experiments were suggested, in order to strengthen 
evidence of differential regulation. However, functional binding 
experiments are experimentally costly at scale, and are typically 
applied to specific systems of interest. As an alternative, 
several \textit{in silico} tools have been developed to predict 
SNP effects on miRNA-gene interactions~\cite{barenboim2010microsniper,deveci2014mrsnp}. 
However, these tools often fail to predict interactions 
that have been been observed in experiment~\cite{chi2012alternative}. 

To date, the functional effects of polymorphisms are typically 
explored by integrating GWAS and gene expression data
find expression Quantitative Trait Loci (eQTLs): SNP variants
that result in altered gene expression. Many eQTLs 
have been identified, including several associated 
with cancer. Recent integrative analyses using data from 
The Cancer Genome Atlas (TCGA) identified eQTLs in 
Breast Cancer~\cite{li2013integrative} and Glioblastoma Multiforme~\cite{chen2014systematic,shpak2014eqtl}. In fact, combinations of GWAS data with eQTL studies 
have found alleles that affect gene expression and 
complex traits genome-wide~\cite{zhu2016integration}. 
However, these analyses do not necessarily reveal the functional 
effects of polymorphisms on molecular-molecular 
interactions, particularly with respect to differential 
binding, as in miRNA-gene or TF-gene interactions. 

Data from the TCGA project provides an ideal opportunity
to investigate the function of genetic vairants by
integrating SNP, gene expression, and miRNA expression
from the same set of samples.
Here, we propose a method to integrate these data
to reveal genetic variants that show evidence of impacting
miRNA-gene regulatory relationships.
Motivated 
by the observation that integrative omics analyses provide 
more insight than single-platform approaches~\cite{kristensen2014principles, SUN2016147}, 
we perform an integrative omics analysis that searches for 
polymorphisms that modulate co-expression between miRNAs and 
their putative gene targets, which we term ``regulatory QTLs (regQTLs)'': loci 
whose alleles impact the regulation of genes by miRNAs. Using mRNA 
expression, miRNA expression, and genotype data taken from tumor 
tissues, our method applies a regression model to assess whether 
disparate alleles present at a genomic variant modulate the 
miRNA-gene co-regulatory relationship.  By comparing miRNA 
expression and gene expression across genotypes, we can
identify regQTLs, or polymorphic sites which may 
alter molecular interactions and may be implicated in 
tumorigenesis. 
Importantly, by using miRNA and gene expression
data, we avoid the inaccuracies associated with miRNA binding
prediction algorithms, and are able to directly estimate the magnitude
of the impact that the SNP has on the regulatory relationship.

Below, we present the method and apply it to TCGA data from
four separate cancer types (breast, lung, liver, prostate).
We report findings of gene variants that modulate miRNA
regulation of gene expression in each of the cancer types studied.
Interestingly, some of the flagged miRNAs and genes have been
previously implicated in tumorigenic processes in the literature, and
SNPs demonstrate functional changes to gene regulation. These results
may have implications for future research in genomic regulation in
tumors.

\section{Methods}

We seek to identify regQTLs, genomic variants that influence miRNA regulation
of gene expression, by integrating genomic and expression data from
TCGA data. Specifically, we  test whether different alleles at a SNP
locus within a given gene alters how a miRNA modulates the
expression of that gene across TCGA tumor samples. regQTLs
may then provide context to gene regulation in cancer, 
due to genetic diversity or genetic alterations.

Previously~\cite{wilk2017braun}, we had identified 
sets of genes, or pathways, whose overall activity appeared 
to be dysregulated by miRNAs in tumors in comparison to 
healthy tissue in four separate cancer types (breast, lung, liver, prostate). 
Our method first obtained an expression-based summary of 
pathway activity using Isomap~\cite{tenenbaum2000global}, 
and then searched for differential miRNA correlations with 
the pathway summary across phenotypes, to find miRNA-pathway 
relationships at the systems level that were disrupted in cancer. 
Using data from The Cancer Genome Atlas (TCGA), we tested ~${\sim}10^5$ 
unique miRNA-pathway relationships, many of which were significantly dyregulated. 

Here we focus on those dysregulated miRNA-pathway pairs, 
and explore whether SNPs on the pathways are responsible for 
miRNA dysregulation of individual genes within that pathway. In other words, for each 
miRNA-pathway pair, we explore the co-expression patterns 
between the miRNA and the genes on the pathway, modulated 
by each of the polymorphisms located on the gene. By restricting 
our focus to genes in dysregulated miRNA-pathway pairs, we can ensure 
that the polymorphisms under consideration reside within perturbed 
systems in cancer. In addition, this restriction effectively reduces the 
dimensionality of our genome-wide analysis. We apply our methodology 
to TCGA data to explore all miRNA-mRNA-SNP combinations from miRNA-gene
pairs where the gene was part of a dysregulated miRNA-pathway system,
amounting to ~${\sim}10^6$ models per cancer type (breast, lung, liver, prostate). 
For each cancer, we report regQTLs which appear to modulate the 
co-regulatory miRNA-gene relationship in tumors and may therefore contribute to tumorigenesis.


\subsection{Analytical approach}
We consider all miRNA-gene pairs from dysregulated pathways that 
exhibited a differential association $p<0.01$ in
our prior analysis~\cite{wilk2017braun}.
We systematically probe all unique miRNA-mRNA-SNP trios across all
tumor samples in a cancer cohort. For each unique trio, we compute a
multiple linear regression to model the expression of a gene
as a response as a function of the miRNA expression,
the SNP allele (treated as a categorical variable),
and the interaction between them. 
%
Specifically, for a SNP with genotypes $\{AA,Aa,aa\}$, we fit
\begin{equation}\label{eq:lm}
\begin{split}
Y = \beta_{0} + \beta_{1}X_{\text{miR}} + \beta_{2}\indi{\text{SNP=Aa}} + \beta_{3}\indi{\text{SNP=aa}} +\\ 
\beta_{4}\indi{\text{SNP=Aa}}X_{\text{miR}} + \beta_{5}\indi{\text{SNP=aa}}X_{\text{miR}} +\varepsilon\,,
\end{split}
\end{equation}
where $Y$ represents the expression level of the gene of interest,
$X_{\text{miR}}$ is the expression level of the miRNA, and
$\indi{\cdot}$ is an indicator function for the SNP genotype.
In this model, the coefficient $\beta_1$ quantifies the
relationship between the miRNA and gene expression for the reference
genotype $AA$; the coefficients $\beta_2, \beta_3$ quantify how the
allele affects overall expression of the gene (\ie, as an eQTL);
and the interaction coefficients $\beta_4, \beta_5$ quantify how the variant alleles
at the SNP of interest modulate the miRNA-mRNA relationship. 
SNPs with strong interaction effects are inferred to be potential 
regQTLs.


SNPs are treated as categorical variables in our model to 
capture any dominant, recessive, or additive effects that individual 
alleles may confer on miRNA-gene interactions. Because any single 
copy of an allele may create, strengthen, weaken, or abrogate 
miRNA-gene binding, we seek to capture all possible SNP effects and 
their cross-comparisons. For instance, an allele that creates
strong miRNA-gene binding may only need to be present in one
copy to show an effect, such that the salient difference is
observed between having no copy of the variant allele and having
one or two copies (with no difference between one and two). 
Alternatively, an allele that abrogates miRNA-gene binding may
be seen to have a strong effect for those with homozygous copies,
but a much weaker effect for heterozygous individuals.
%
As such, we explore all allelic effects on 
miRNA-gene binding.

\begin{figure*}[htb]
\begin{center}
\includegraphics[scale=0.5]{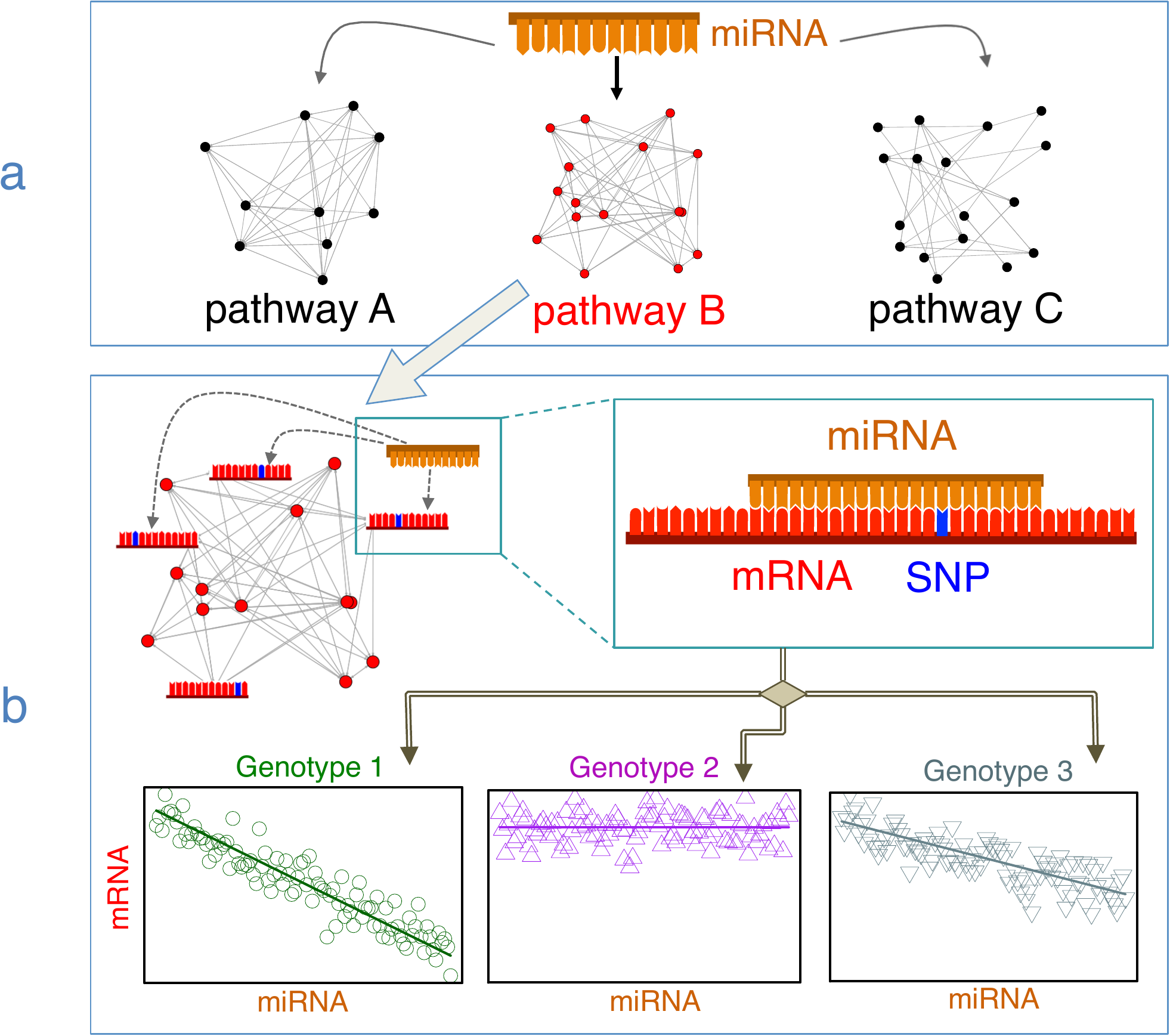}
\end{center}
\caption{Illustration of the procedure to identify regQTLs: SNPs that modify the miRNA-mRNA relationship in dysregulated pathways. The method integrates gene expression, miRNA expression, and SNP data. 
(a) To aid mechanistic interpretability and reduce the search space, we first identify miRNA--pathway pairs that exhibit significant evidence of differential regulation following~\cite{wilk2017braun}. (b) Within each miRNA--identified pathway pair, we construct all miRNA-mRNA-SNP trios for each gene in the pathway (top), and systematically test whether the SNP modifies the expression relationship between the miRNA and the mRNA (bottom).  Table~\ref{tab:procedure} details the method.}     
\label{fig:flowchart} 
\end{figure*}

To assess the statistical significance of the interaction effect,
we apply ANOVA Type III sums of squares (Yates's
weighted squares of means) to compare the full model to that without
the interaction terms. A significant $F$ statistic for the interactions suggests
that at least one of the variant  SNP alleles substantially alters the relationship  between
the miRNA and the mRNA, on top of any eQTL-like effects. 
$p$-values for all interactions are then FDR-adjusted~\cite{benjamini1995controlling} for the
large number of miRNA-mRNA-SNP trios probed in the dataset.
(We choose this Benjamini-Hochberg FDR adjustment  because we expect that models with common
miRNAs or genes are not strictly independent, and this procedure has been show to provide control of the
FDR under dependency~\cite{benjamini2001control}.)
The steps of the method are summarized in Table~\ref{tab:procedure}. Figure~\ref{fig:flowchart} illustrates the intuition underlying the method. 

\subsection{Application to TCGA data}
As a proof of concept, we applied this method systematically to tumor
samples with combined miRNA expression, gene expression, and SNP
genotype data from TCGA.

\paragraph{Data}
TCGA data were downloaded for BRCA (breast), LIHC (liver), LUSC
(lung), and PRAD (prostate) cancers. Tumor samples (TCGA sample type
``01'') measured across mRNA IlluminaHiSeq\_RNASeqV2 (Level 3), miRNA
IlluminaHiSeq\_miRNASeq (Level 3), and Affymetrix SNP6.0 platforms
were used for the analysis, amounting to 699 total tumor samples in
breast, 345 in liver, 341 in lung, and 481 in prostate cancer.

\paragraph{Data preprocessing and filtration}
Briefly, mRNA data were converted to TPM and log2 transformed, and
miRNA data were log2 transformed (both with small offsets for the log
transformation). In addition, genes and miRNAs were removed from
consideration that had very low expression across most samples in the
set (defined as genes with median expression $<10^{-9}$ before
TPM conversion and miRNAs having expression $\leq 1$ for more than
half of the samples in the set before log transformation).

SNPs were filtered out that had a low Birdseed confidence threshold
(0.05) for genotype calls in the TCGA pipeline. We used two additional
filtration criteria to remove SNPs: a) those having minor allele
frequencies (MAF) less than 1\% and b) those having genotype
frequencies less than 5\% across all samples in a cancer dataset.
These criteria were imposed to ensure that limited sampling of
rare alleles and genotypes would not skew the regression results.

Before applying the regression models, individual samples within a
miRNA-mRNA-SNP trio having no appreciable miRNA expression
were removed from consideration, since they are not biologically of
interest. Additionally, samples having a large Cook's Distance ($D >
1$) were removed from the regressions and the regressions were
recomputed to limit the influence of outliers on the resulting models.

\begin{table}
\tableparts{%
\caption{Procedure for assessing genomic variants modulating miRNA-gene interactions}
\label{tab:procedure}
}{%
\begin{tabular*}{\columnwidth}{r@{.~}p{0.95\columnwidth}}
\toprule
\multicolumn{2}{c}{\textbf{Method for finding regQTLs}} \\
\colrule
1& Select dysregulated miRNA-pathway pairs ($p<0.01$) following the method from \cite{wilk2017braun} (Figure~\ref{fig:flowchart}a). \\
2& For each miRNA-pathway pair, find all genes on the pathway and all assayed SNPs on each gene to construct all unique miRNA-mRNA-SNP trios (Figure~\ref{fig:flowchart}b, top). \\
3& For each trio in Step 2, fit Equation~\ref{eq:lm} 
and apply ANOVA to assess statistical significance of the interaction terms (Figure~\ref{fig:flowchart}b, bottom).\\
4& FDR-adjust the resulting ANOVA $p$-values. \\
5& Report highly significant miRNA-mRNA-SNP trios as potential regQTLs.\\
\botrule
\end{tabular*}
}{%
} 
\end{table} 

\section{Results}
\begin{figure*} 
\begin{center}
\includegraphics[width=\textwidth]{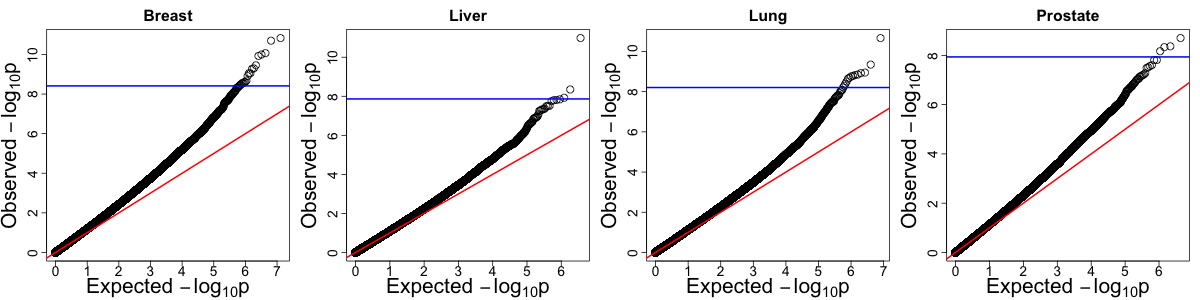}
\end{center} 
\caption{Quantile-quantile plots of the observed $p$-values for the
gene-miRNA-SNP ANOVA interaction tests versus their expected $p$-value
distributions (the uniform distribution), tested in each cancer type.
There were approximately $1.29\times10^7$ unique interactions tested
in breast, $3.65\times10^6$ in liver, $8.03\times10^6$ in lung, and
$4.32\times10^6$ in prostate cancer. A horizontal blue line indicates the
threshold for genome--wide significance under the conservative 
Bonferroni adjustment.
}
\label{fig:qq} 
\end{figure*}

We begin by presenting $qq$-plots of the regQTL $p$-values across
all miRNA-mRNA-SNP trios in TCGA breast, liver, lung, and prostate
cancer samples (Figure~\ref{fig:qq}).  It can be seen here that 
several trios in each study achieve extremely small $p$-values of $p\leq10^{-9}$,
indicating regQTLs that achieve genome--wide significance
(even using the conservative Bonferroni correction).

It may also be observed that the distribution of regQTL $p$-values exhibit systematic 
deviations from the expected uniform distribution of $p$ values under the null, with many more significant observations
than expected by chance for independent tests (as demonstrated by the trend away from the red
diagonal lines). 
Such systematic deviations suggest that the trios are not strictly
independent of one another; in classical GWAS, this is often 
attributable to population substructure driving the results.
Here, however, some dependency amongst the tests is 
expected. Because we consider 
all known SNPs on each gene, many of the SNPs will be in linkage disequilibrium (LD) 
owing to their genomic proximity and will be correlated.  Variants in LD have 
been observed in blocks ranging from tens of Kbs to greater 100 Kbp~\cite{reich2001linkage}, 
which may be larger than the size of a gene. In addition, because we consider 
genes within pathways, the expression of the genes may be correlated due 
to similar co-regulatory mechanisms or cooperate effects within a network. 
Population substructure may also be a factor in data drawn from diverse 
genetic populations. However, TCGA heavily samples from European ancestry;
we tested for substructure by applying PCA to genotype data, and 
found that most of the samples comprised a single tight cluster in the first two principal 
components, as shown in Supplementary Figures S1--S4. 
Because we expect the tests to exhibit some dependency,
we perform multiple hypothesis adjustment using FDR~\cite{benjamini1995controlling,benjamini2001control},
rather than using the Bonferroni adjustment, which assumes independent
tests and can be excessively conservative otherwise.

\subsection{Breast cancer}

\begin{figure*}[htb]
\begin{center}
\includegraphics[scale=0.37]{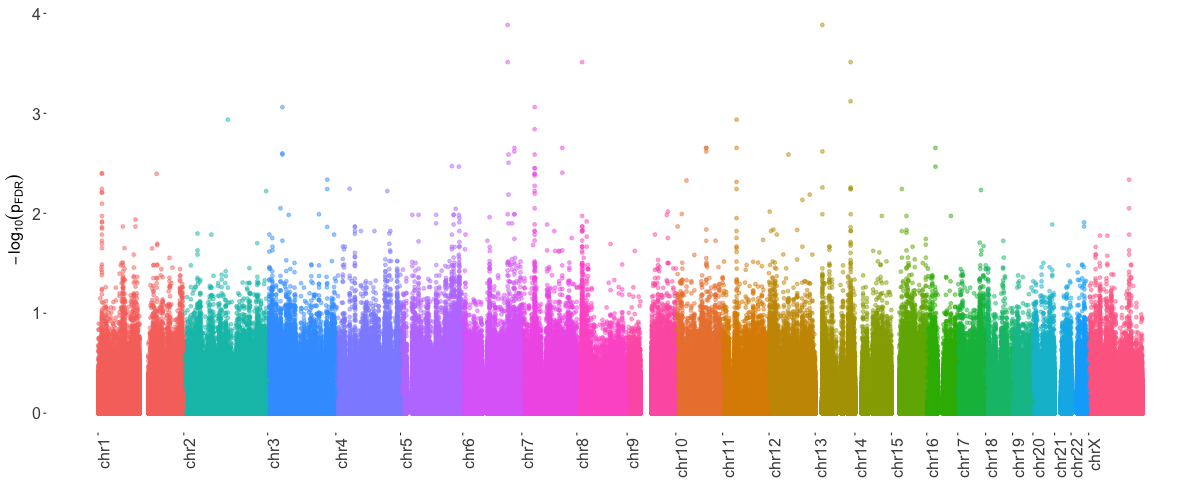}
\end{center}
\caption{Breast cancer manhattan plot. All gene-miRNA-SNP interaction
$p$-values mapped to the location of the SNP in the genome.
Observations are colored by chromosome. $p$-values are adjusted for
the False Discovery Rate.} 
\label{fig:BRCAmanhattan} 
\end{figure*}

In breast cancer, ${\sim}1.28 \times 10^7$ unique
gene-miRNA-SNP trios, drawn from 25,850 miRNA$\times$pathway pairs,
were analyzed and shown in Figure~\ref{fig:BRCAmanhattan}. Several
chromosomes contain clusters of significant observations, as
demonstrated by upward spikes within specific genomic regions. These
clusters are composed of SNPs at different loci in close proximity
with one another whose alleles are in
LD. SNPs in LD are influenced by rates of
recombination and mutation and reflect evolutionary history.
Because of their genomic proximity, SNPs in linkage often lie within
the same genes, such that multiple variants in a gene may wield
similar biological effects on miRNA regulation, as demonstrated in
Figure~\ref{fig:BRCAmanhattan}.

This is evident when we observe regQTLs, all trios that were flagged as
significant after FDR-adjustment ($FDR\leq0.1$), of which
relatively few achieve significance (3369). Among the flagged
regQTLs, some miRNA-gene pairs are represented frequently,
with multiple SNPs in the gene appearing to modulate the miRNA-gene relationship.
Table~\ref{tab:BRCAtable} lists the miRNA-gene pairs with the largest
number of significant regQTLs achieving significance with $FDR\leq0.1$.
(Note that Table~\ref{tab:BRCAtable} is not an
exhaustive list but rather displays the highest number of modulating 
SNPs within miRNA-gene interactions at a given FDR.)

\begin{table}[htb]
\tableparts{%
\caption{miRNA-gene
pairs containing the most number of gene variants significantly
modulating their interactions in breast cancer. ``SNPs''
indicates the number of associated SNPs on the gene found to significantly
modulate (at $p_{FDR}\leq0.1$) the miRNA-gene interaction, out of the
total number of known SNPs on the gene.  $p_{MIN}$ indicates the
most significant interaction $p$-value after FDR-correction. ``chr''
indicates the chromosome where the SNP is located.``target'' indicates
whether the miRNA is predicted to target the gene based off sequence
matching from microRNA.org.} 
\label{tab:BRCAtable}
}{%
\begin{tabular*}{\columnwidth}{l@{~}l@{~}c@{~}clcl}
\toprule
 \multicolumn{7}{c}{SNPs} \\ 
 \cline{3-4}
  miRNA & gene & associated & total & $p_{MIN}$ & chr & target \\
  \colrule
  hsa-mir-221 & \textit{FGF14} & 14 & 334 & 7.57E-04 & 13 & TRUE \\
  hsa-mir-642a & \textit{MTOR} & 14 & 38 & 3.97E-03 & 1 & FALSE \\ hsa-mir-141
  & \textit{FOXO1} & 13 & 44 & 2.16E-02 & 13 & TRUE \\ hsa-mir-642a & \textit{NUP210} &
  13 & 173 & 4.93E-02 & 3 & FALSE \\ hsa-mir-154 & \textit{ELMO1} & 10 & 313 &
  8.67E-04 & 7 & FALSE \\ hsa-mir-200c & \textit{FOXO1} & 10 & 44 & 4.13E-02 &
  13 & TRUE \\ hsa-mir-200c & \textit{GALNT10} & 10 & 123 & 1.04E-02 & 5 & TRUE
  \\ hsa-mir-3200 & \textit{GSPT2} & 9 & 15 & 4.46E-02 & X & FALSE \\
  hsa-mir-125b-2 & \textit{LRRC4C} & 9 & 479 & 4.85E-03 & 11 & FALSE \\
  hsa-mir-141 & \textit{EDA2R} & 8 & 37 & 6.66E-02 & X & TRUE \\ hsa-mir-222 &
  \textit{FGF14} & 8 & 334 & 3.07E-04 & 13 & TRUE \\ hsa-mir-141 & \textit{SFRP2} & 8 &
  64 & 1.59E-02 & 4 & FALSE \\ hsa-mir-190b & \textit{TUSC3} & 8 & 235 &
  3.07E-04 & 8 & TRUE \\ hsa-mir-452 & \textit{TUSC3} & 8 & 235 & 1.50E-02 & 8
  & FALSE \\ hsa-mir-190b & \textit{NUP210} & 7 & 173 & 1.77E-02 & 3 & FALSE \\
  \botrule
  \end{tabular*}
 }{%
 }
\end{table}

It is notable that in several pairs in Table~\ref{tab:BRCAtable}, the
miRNA is not predicted to target the gene based off sequence matching
from microRNA.org~\cite{betel2008microrna}. This may be due to several
factors. First, not all miRNA-gene interactions are known, and some
interactions have been observed in experiment that have not been
predicted through sequence matching~\cite{chi2012alternative}. Because
these miRNA-gene pairs are modulated by many gene variants, these
may represent novel biological interactions
between miRNAs and genes that have yet to be documented and that are
sensitive to biological variation. Another possibility is that these
miRNAs and genes may not interact directly, but may be indirectly
connected through second-order effects---for instance, one can
envision a miRNA \textit{target} interacting with the gene listed in the pair
in Table~\ref{tab:BRCAtable}. This may lead to an apparent association
with the miRNA, although it is mediated through another gene.

\begin{figure*} 
\centering
\includegraphics[width=\textwidth]{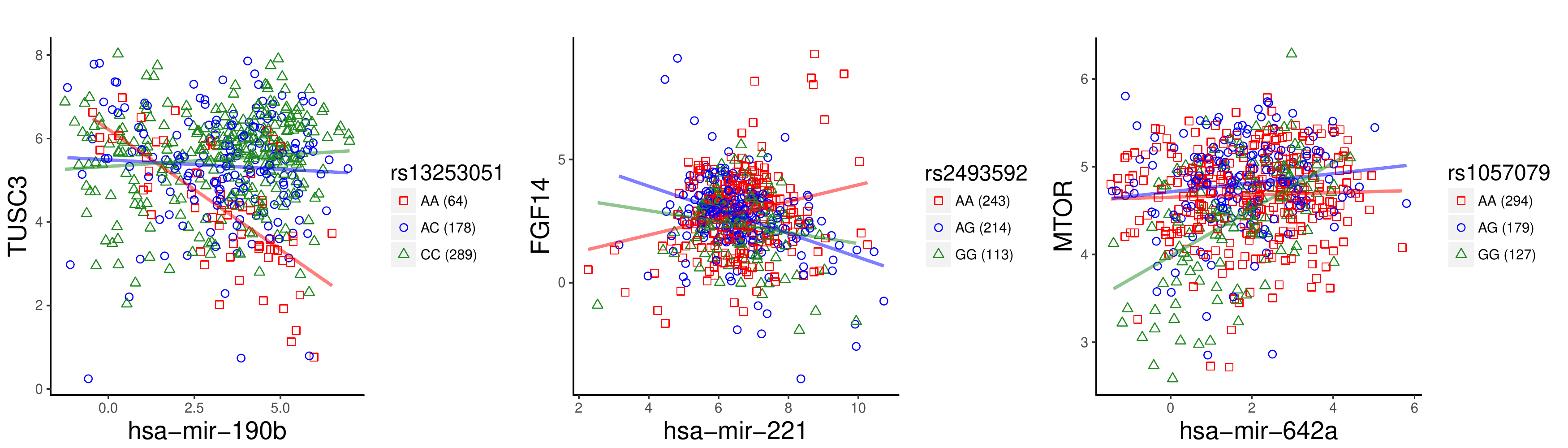}
\caption{Examples of breast cancer gene-miRNA-SNP trios with
significant regulatory differences across genotypes. All trio
interactions $p_{FDR}<0.005$.} 
\label{fig:BRCAtrios} 
\end{figure*}

Examples of significant trios are shown in
Figure~\ref{fig:BRCAtrios}. For instance, in the left panel, samples with the
homozygous minor (AA) genotype exhibit a strong negative dependence
between hsa-mir-190b and \textit{TUSC3}, whereas the heterozygous (AC) and
homozygous major (CC) genotypes exhibit weaker and no dependencies.
One explanation may be that samples having both A alleles confer
strong binding between hsa-mir-190b and \textit{TUSC3}, whereas the
introduction of the C allele confers weaker (AC) or no binding (CC) at
all. hsa-mir-190b is predicted to target \textit{TUSC3} by sequence matching,
and both the miRNA and gene are implicated in cancer in the
literature. \textit{TUSC3} is a tumor suppressor whose loss or decreased
expression is associated with the proliferation of several cancer
types~\cite{vavnhara2013loss,horak2014tusc3,fan2016decreased} and is
markedly under-expressed in breast cancer
cells~\cite{poola2009molecular}. hsa-mir-190b has recently been found
to be the most upregulated miRNA in ER$\alpha$ breast cancers relative
to ER$\alpha$ negative breast cancers~\cite{cizeron2015m}, and is part
of the regulatory network that activates p53~\cite{yu2014nf}. Likewise,
in the middle panel, hsa-mir-221 is predicted to target \textit{FGF14} and
exhibits regulatory differences across genotypes. In this case, the
homozygous minor (AA) appears to confer a loss of regulation, whereas
the introduction of the G allele in the heterozygous (AG) and
homozygous major (GG) genotypes, confers negative regulation and
perhaps strong binding. hsa-mir-221/222 has previously been associated
with a basal-like phenotype and the epithelial to mesenchymal
transition in breast cancer~\cite{stinson2011trps1}. Although \textit{FGF14} in
particular is not implicated in cancer, aberrant signaling of other
Fibroblast Growth Factors are widely found in the pathogenesis of
cancer~\cite{turner2010fibroblast}.

In the right panel, the anomalous genotype, homozygous major (GG),
is associate with a strong positive dependence between hsa-mir-642a and \textit{MTOR},
in contrast to the other panels. The positive dependence, coupled with
the fact that hsa-mir-642a is not predicted to target \textit{MTOR}, may be
evidence that our method is flagging second-order effects driven by
genomic variation. Furthermore, hsa-mir-642, the miRNA family of
hsa-mir-642a, contains many genes that interact with \textit{MTOR}, including
many MAP kinases and \textit{TP53} that are implicated in breast cancer in the
literature. The examples shown in Figure~\ref{fig:BRCAtrios} are
illustrative of the types of regulatory interactions affected by
genomic variation we observe within breast cancer.

\subsection{Liver cancer}

\begin{figure*}[htb] 
\centering
\includegraphics[scale=0.37]{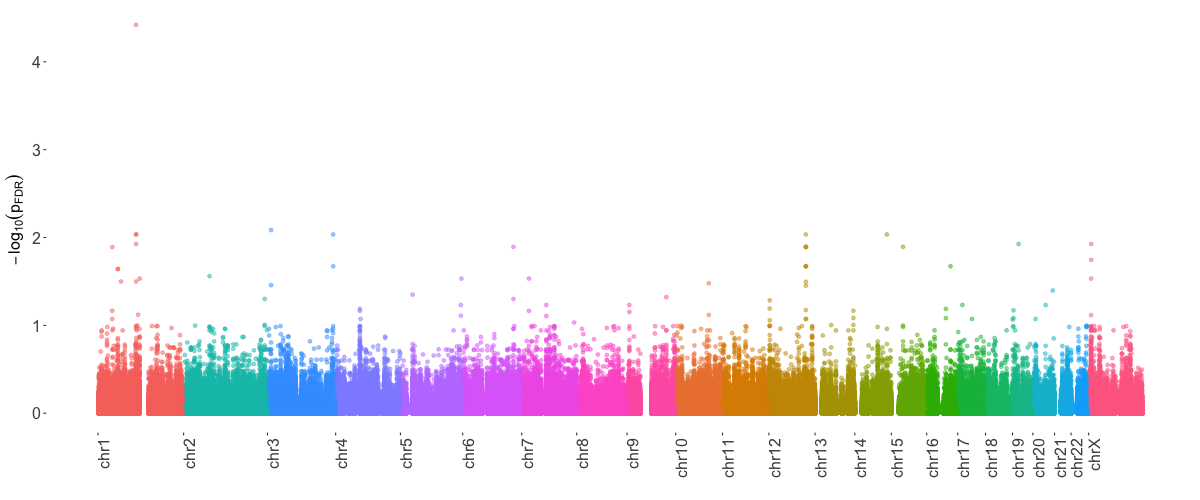} 
\caption{Liver cancer manhattan plot of regQTL $-log_{10}FDR$ values.} 
\label{fig:LIHCmanhattan}
\end{figure*}

\begin{figure}[htb] 
\begin{center}
\includegraphics[width=\linewidth]{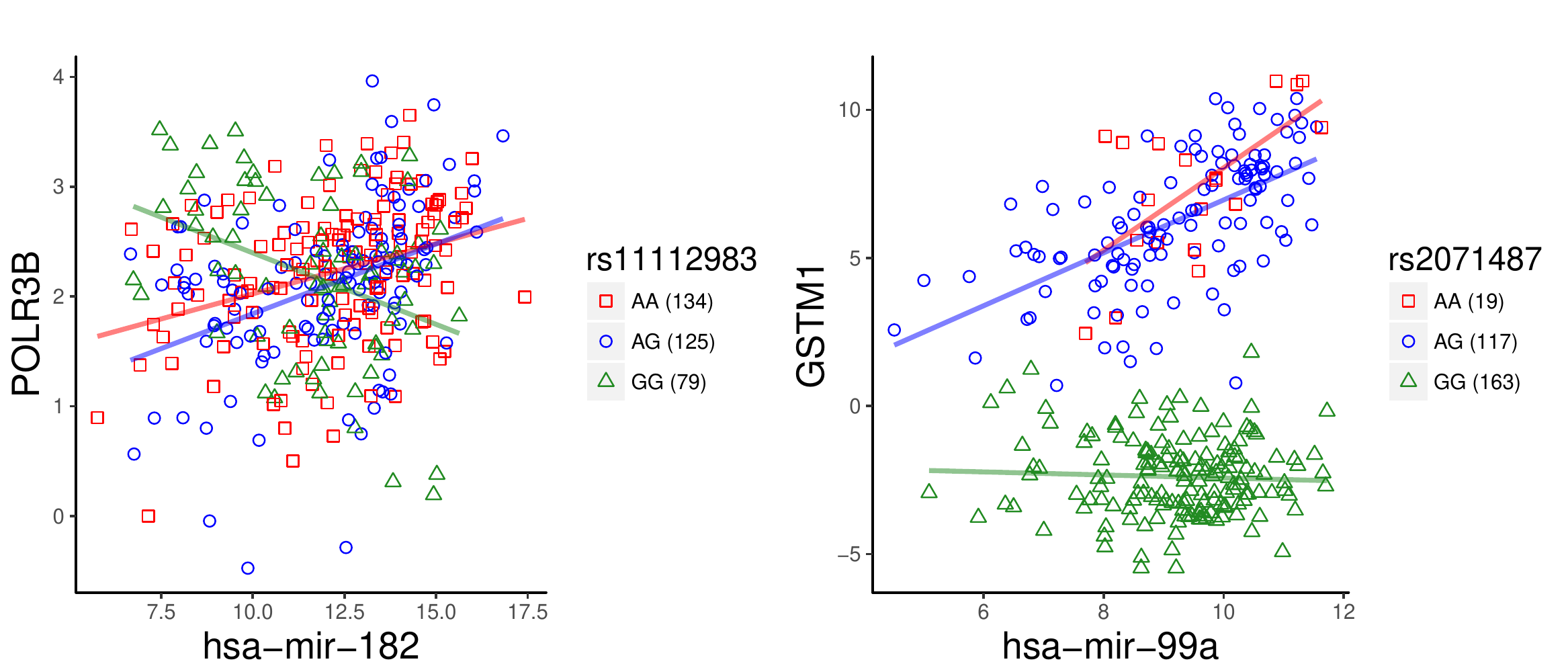} 
\end{center}
\caption{Examples of liver cancer gene-miRNA-SNP trios with
significant regulatory differences across genotypes. All trio
interactions $p_{FDR}<0.005$.} 
\label{fig:LIHCtrios} 
\end{figure}

A total of
$3.65\times10^6$ gene-miRNA-SNP unique trios, drawn from 7371
miRNA$\times$pathway pairs, were mapped to their loci in the genome in
Figure~\ref{fig:LIHCmanhattan} for liver cancer. 
Although fewer trios achieve significance
in comparison to breast cancer (76 regQTLs with $FDR\leq0.1$), we again
observe the associated SNPs to be clustered due to linkage.  
As above, Table~\ref{tab:LIHCtable} presents the miRNA-gene pairs
that have the greatest number of regQTLs.
%
The top gene,
\textit{POLR3B}, contains 6 SNPs that each modulate its regulation by
hsa-mir-182, depending on genotype. We illustrate one such example in
Figure~\ref{fig:LIHCtrios} (left plot), in which the anomalous genotype
(GG) for rs11112983 downregulates expression of \textit{POLR3B} by hsa-mir-182,
in comparison to the others. \textit{POLR3B} is subunit B (the second largest)
of RNA polymerase III, which is the polymerase that synthesizes
transfer and small ribosomal RNAs. Increased RNA polymerase III output
is widely implicated in cancer~\cite{marshall2008non}. Recently, a
novel truncated version of \textit{POLR3B} called \textit{INMAP} has been observed to
repress \textit{AP-1} and \textit{p53} activity and is upregulated in several cancer
cell lines~\cite{yunlei2013inmap}. hsa-mir-182 is significantly
upregulated in hepatocellular carcinoma and has been found to promote
proliferation and invasion by downregulating tumor suppressor
\textit{EFNA5}~\cite{wang2016oncomir} and promote metastasis by
downregulating metastasis suppressor 1~\cite{wang2012microrna}. While
hsa-mir-182 itself isn't predicted to target \textit{POLR3B}, it is predicted
to target other subunits on RNA polymerase III, and therefore may
exert second-order regulatory effects with \textit{POLR3B}.

Another example of noteworthy genotype-dependent interactions we
detect is shown in Figure~\ref{fig:LIHCtrios} (right plot). Not only is
\textit{GSTM1} differentially regulated by hsa-mir-99a at rs2071487 depending
on genotype, but also \textit{GSTM1} exhibits initial genotype-dependent gene
expression differences typical of a strong eQTL. In this case, alleles appear to have the power
to determine both expression and modulation of gene regulation. \textit{GSTM1}
is part of the GST-superfamily that detoxifies electrophilic compounds
by conjugation with glutathione, and is involved in processing
carcinogens, drugs, and toxins. \textit{GSMT1} is highly polymorphic, affecting
toxicity and drug efficacy across individuals, and in particular, null
mutations are associated with an increase in susceptibility to lung,
bladder, and colon cancers~\cite{zhong1993relationship}. hsa-mir-99a
inhibits hepatocellular carcinoma growth~\cite{li2011microrna} and its
dysregulation is an early marker of tumor
progression~\cite{petrelli2012sequential}.  While hsa-mir-99a is not
predicted to target \textit{GSTM1}, it is predicted to target \textit{GSTM3} and \textit{GSTM5},
other GST-superfamily $\mu$ enzymes.

\begin{table}[htb]
\tableparts{%
\caption{miRNA-gene pairs with the greatest number of significant regQTLs (at
$p_{FDR}\leq0.1$) in liver cancer.}
\label{tab:LIHCtable}
}{%
\begin{tabular*}{\columnwidth}{l@{~}l@{~}c@{~}clcl}
\toprule
\multicolumn{7}{c}{SNPs} \\
   \cline{3-4}
   miRNA & gene & associated & total & $p_{MIN}$ & chr & target \\
   \colrule
   hsa-mir-182 & \textit{POLR3B} & 6 & 34 & 9.22E-03 & 12 & FALSE \\
  hsa-mir-183 & \textit{POLR3B} & 6 & 34 & 1.28E-02 & 12 & FALSE \\ hsa-mir-107
  & \textit{NFYC} & 3 & 21 & 1.28E-02 & 1 & TRUE \\ hsa-mir-125b-1 & \textit{STS} & 3 &
  212 & 1.18E-02 & X & TRUE \\ hsa-mir-122 & \textit{ATP2B2} & 2 & 237 &
  8.21E-03 & 3 & TRUE \\ hsa-mir-655 & \textit{CACNA1C} & 2 & 301 & 5.19E-02 &
  12 & TRUE \\ hsa-mir-215 & \textit{DRD1} & 2 & 84 & 5.87E-02 & 5 & FALSE \\
  hsa-mir-139 & \textit{GLDC} & 2 & 51 & 5.87E-02 & 9 & FALSE \\ hsa-mir-203 &
  \textit{PLCE1} & 2 & 118 & 3.32E-02 & 10 & FALSE \\ hsa-mir-766 & \textit{THOP1} & 2 &
  3 & 6.74E-02 & 19 & TRUE \\ hsa-mir-34c & \textit{UGT1A6} & 2 & 43 & 5.01E-02
  & 2 & FALSE \\
  \botrule
  \end{tabular*}}
 {}
 \end{table}

\subsection{Lung cancer}

\begin{figure*}[htb]
\begin{center}
\includegraphics[scale=0.37]{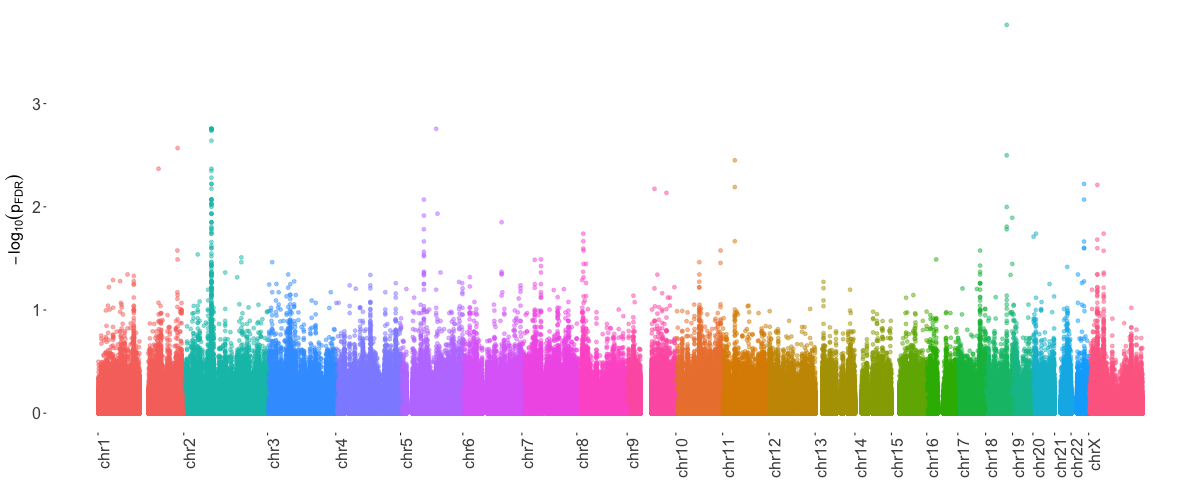} 
\end{center}
\caption{Lung cancer manhattan plot of regQTL $-log_{10}FDR$ values.} 
\label{fig:LUSCmanhattan}
\end{figure*}

A total of 
$8.03\times10^6$ unique gene-miRNA-SNP trios, drawn from from 14433
miRNA$\times$pathway pairs in lung cance were mapped to their loci
in the genome in Figure~\ref{fig:LUSCmanhattan}. Among the clusters of
correlated observations that spike in Figure~\ref{fig:LUSCmanhattan},
several miRNA-gene pairs are represented frequently and tabulated in
Table~\ref{tab:LUSCtable}. \textit{MAOA}, \textit{POLA1}, and \textit{CTNNA2} on Chromosomes X
and 2 collectively make up the bulk of the genes containing SNPs
modulating their observations in LD with each other.

\begin{figure}[htb]
\begin{center}
\includegraphics[width=\linewidth]{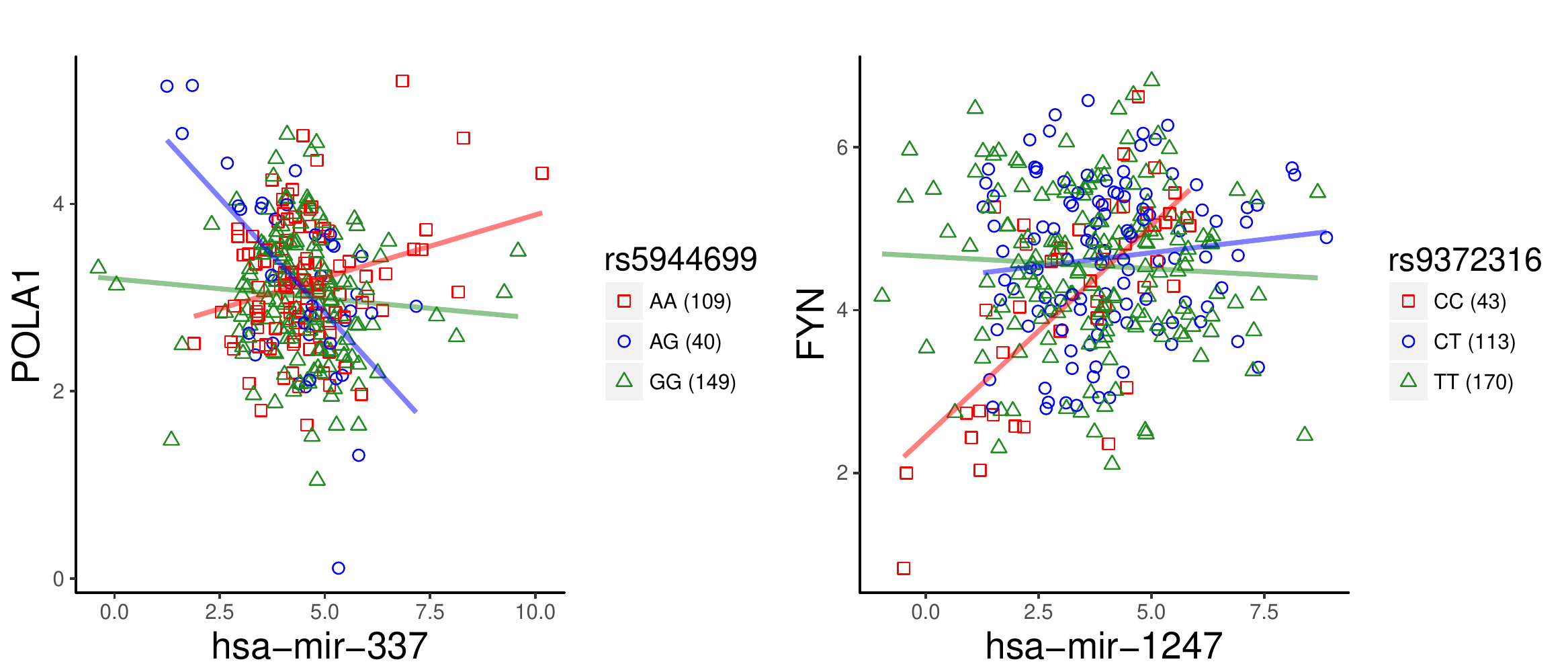} 
\end{center}
\caption{Examples of lung cancer gene-miRNA-SNP trio with a
significant regulatory difference across genotypes. All trio
interactions $p_{FDR}<0.015$.} 
\label{fig:LUSCtrios} 
\end{figure}

In Figure~\ref{fig:LUSCtrios} (left plot), one of the SNPs (rs5944699)
modulating \textit{POLA1} and hsa-mir-337 is shown. All three genotypes exhibit
disparate regulation on the gene. Interestingly, the heterozygous
genotype (AG) demonstrates a negative regulatory effect rather than
the homozygous genotypes. \textit{POLA1} is the catalytic subunit of DNA
Polymerase Alpha 1. hsa-mir-337 regulates the proliferation of several
cancer types, including pancreatic and gastric cancers.

Yet another example of anomalous regulation we detect is shown in
Figure~\ref{fig:LUSCtrios} (right plot). Here the CC genotype for
rs9372316 exhibits a strong positive correlation between hsa-mir-1247
and \textit{FYN}, in contrast to the other genotypes, which exhibit no
appreciable correlation. hsa-mir-1247 is not predicted to target \textit{FYN},
providing additional evidence that we may be detecting secondary
effects of allele-specific differences on miRNA regulation of genes.
hsa-mir-1247 is predicted to target several genes that interact with
\textit{FYN}, including \textit{TNFRSF10B} (tumor necrosis factor receptor) and \textit{CBL}
(proto-oncogene), that are implicated in tumorigenic processes. \textit{FYN}
itself is a proto-oncogene of the Src family normally associated with
T-cell signaling, cell development and growth. \textit{FYN} is upregulated in
several cancer types including breast and prostate
cancers~\cite{posadas2009fyn} and is associated with metastatic
potential.

We note that in Table~\ref{tab:LUSCtable}, \textit{CTNNA2}
appears frequently with several miRNAs, often with the same SNPs.
Because the anomalous SNPs in \textit{CTNNA2} have relatively low
representation among the cohort, it is difficult to attribute
confidence to their regulatory effect. That is, genotype frequencies for the 
homozygous minor alleles (MAF: 28\%--41\%) may be undersampled for 
these SNPs (range: 9.9\%--16.1\%). Another note is that 
\textit{CTNNA2} has low gene expression in lung tumor samples 
($<0.1$ TPM for $\sim$~92\% of the samples). This may subject it
to biological fluctuations, due to noise and excluded samples,
which may influence results.

\begin{table}[htb]
\tableparts{%
\caption{miRNA-gene pairs with the greatest number of significant regQTLs (at
$p_{FDR}\leq0.1$) in lung cancer.}
\label{tab:LUSCtable}
}{%
\begin{tabular*}{\columnwidth}{l@{~}l@{~}c@{~}clcl}
\toprule
\multicolumn{7}{c}{SNPs} \\
   \cline{3-4}
   miRNA & gene & associated & total & $p_{MIN}$ & chr & target \\
   \colrule
   hsa-mir-766 & \textit{MAOA} & 10 & 91 & 1.82E-02 & X & TRUE \\
  hsa-mir-337 & \textit{POLA1} & 8 & 61 & 6.13E-03 & X & TRUE \\ hsa-mir-200b &
  \textit{CTNNA3} & 7 & 654 & 3.44E-02 & 10 & FALSE \\ hsa-mir-127 & \textit{CTNNA2} & 6
  & 732 & 1.76E-03 & 2 & FALSE \\ hsa-mir-134 & \textit{CTNNA2} & 6 & 732 &
  1.76E-03 & 2 & FALSE \\ hsa-mir-154 & \textit{CTNNA2} & 6 & 732 & 8.51E-03 &
  2 & TRUE \\ hsa-mir-369 & \textit{CTNNA2} & 6 & 732 & 4.27E-03 & 2 & TRUE \\
  hsa-mir-379 & \textit{CTNNA2} & 6 & 732 & 6.68E-03 & 2 & FALSE \\ hsa-mir-409
  & \textit{CTNNA2} & 6 & 732 & 1.76E-03 & 2 & TRUE \\ hsa-mir-493 & \textit{CTNNA2} & 6
  & 732 & 8.51E-03 & 2 & FALSE \\ hsa-mir-496 & \textit{CTNNA2} & 6 & 732 &
  8.51E-03 & 2 & FALSE \\ hsa-mir-758 & \textit{CTNNA2} & 6 & 732 & 1.65E-02 &
  2 & FALSE \\ hsa-mir-185 & \textit{ADH4} & 5 & 16 & 4.57E-02 & 4 & TRUE \\
  hsa-mir-382 & \textit{CTNNA2} & 5 & 732 & 3.44E-02 & 2 & FALSE \\ hsa-mir-205
  & \textit{EGFR} & 5 & 125 & 3.23E-02 & 7 & TRUE \\
\botrule
\end{tabular*}}
{}
\end{table}

\subsection{Prostate cancer}

\begin{figure*}[htb] 
\begin{center}
\includegraphics[scale=0.37]{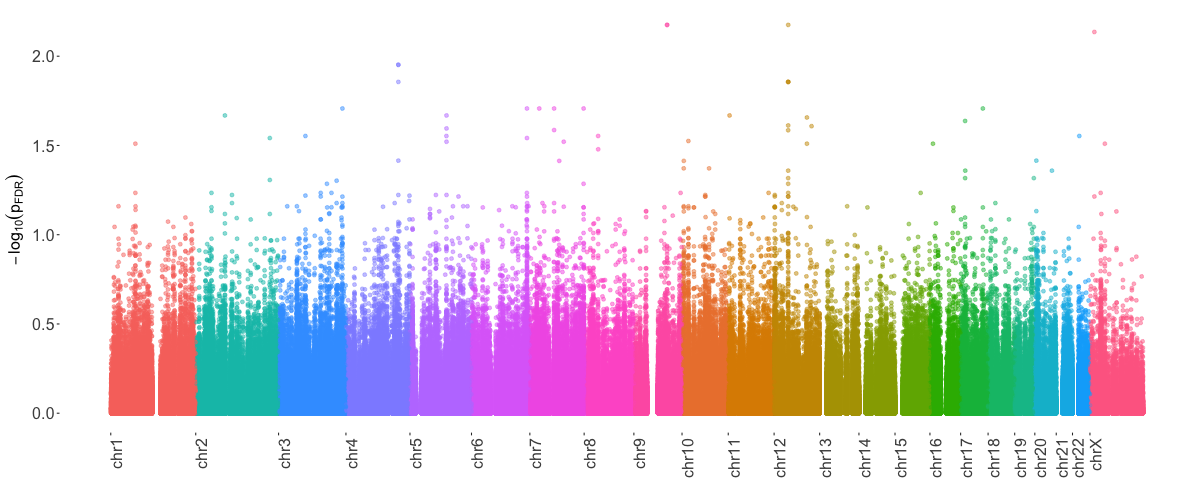} 
\end{center}
\caption{Prostate cancer manhattan plot of regQTL $-log_{10}FDR$ values.} 
\label{fig:PRADmanhattan}
\end{figure*}

A total of
$4.3\times10^6$ unique gene-miRNA-SNP trios, drawn from 8673
miRNA$\times$pathway pairs in prostate cancer were mapped to their
loci in the genome in Figure~\ref{fig:PRADmanhattan}. The scale in
Figure~\ref{fig:PRADmanhattan} is lower than the manhattan plots in the
other cancer types, with the most significant interaction
$p\sim0.0067$. Nevertheless, we do observe pairs containing many SNPs
modulating their interactions in Table~\ref{tab:PRADtable}. In
addition, many of these miRNA-gene pairs are predicted to interact
biologically based on sequence matching.

\begin{figure}[htb]
\begin{center}
\includegraphics[width=\linewidth]{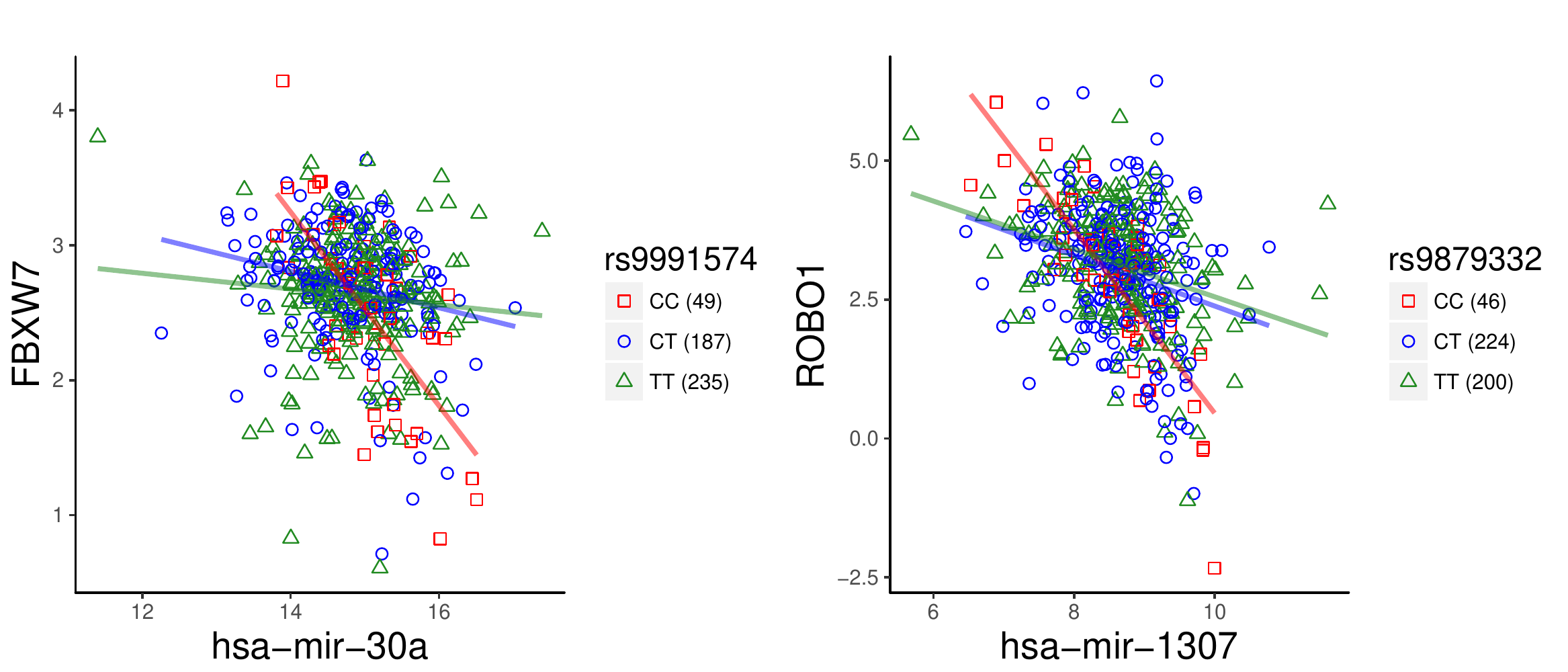} 
\end{center}
\caption{Examples of prostate cancer gene-miRNA-SNP trios with
significant regulatory differences across genotypes. All trio
interactions $p_{FDR}<0.05$.} 
\label{fig:PRADtrios} 
\end{figure}

A few examples of trios we find in prostate cancer are shown in
Figure~\ref{fig:PRADtrios}. In the left plot, hsa-mir-30a is predicted
to target \textit{FBXW7}, and shows a sharp negative regulatory dependence for
the CC genotype, whereas the other genotypes exhibit no appreciable
miRNA-gene regulation. hsa-mir-30a and \textit{FBXW7} are among the most
frequently flagged pairs we find in prostate cancer. hsa-mir-30a is a
tumor suppressor that inhibits EMT genes that is typically
down-regulated by oncogenic signals in prostate cancer like \textit{EGF},
particularly in metastasis~\cite{kao2014mir}. \textit{FBXW7}, an F-box protein,
mediates ubiquitination and proteasomal degradation of target
proteins. Its down-expression, loss, and frequent mutation is shown in
multiple cancer types, including ovarian, breast, melanoma, colon, and
others. In the right plot, the CC genotype exhibits strong negative
regulation between hsa-mir-1307 and \textit{ROBO1}, whereas the other genotypes
exhibit weaker dependencies. hsa-mir-1307 is predicted to target
\textit{ROBO1}, and promotes proliferation in prostate cancer by targeting
\textit{FOXO3A}~\cite{qiu2017mir}. \textit{ROBO1} itself is part of the immonoglobulin
gene superfamily and is an axon guidance receptor gene previously
implicated in dyslexia.

\begin{table}[htb]
\tableparts{%
\caption{miRNA-gene pairs with the greatest number of significant regQTLs (at
$p_{FDR}\leq0.1$) in prostate cancer.}
\label{tab:PRADtable}
}{%
\begin{tabular*}{\columnwidth}{l@{~}l@{~}c@{~}clcl}
\toprule
\multicolumn{7}{c}{SNPs} \\
   \cline{3-4}
   miRNA & gene & associated & total & $p_{MIN}$ & chr & target \\
   \colrule
   hsa-mir-26a-2 & \textit{CNTN1} & 14 & 135 & 6.68E-03 & 12 & TRUE \\
  hsa-mir-15b & \textit{PARK2} & 6 & 636 & 1.96E-02 & 6 & TRUE \\ hsa-mir-30a &
  \textit{FBXW7} & 5 & 33 & 1.12E-02 & 4 & TRUE \\ hsa-mir-1266 & \textit{PDE4D} & 5 &
  534 & 7.17E-02 & 5 & TRUE \\ hsa-mir-143 & \textit{EFNA5} & 4 & 247 &
  2.54E-02 & 5 & FALSE \\ hsa-mir-200c & \textit{FBXW7} & 4 & 33 & 5.99E-02 & 4
  & TRUE \\ hsa-mir-330 & \textit{NEGR1} & 4 & 241 & 6.93E-02 & 1 & TRUE \\
  hsa-mir-421 & \textit{WBSCR17} & 4 & 336 & 1.96E-02 & 7 & FALSE \\
  hsa-mir-130b & \textit{AKR1C3} & 3 & 42 & 3.86E-02 & 10 & TRUE \\ hsa-mir-331
  & \textit{CACNA2D4} & 3 & 53 & 5.99E-02 & 12 & TRUE \\ hsa-mir-766 & \textit{CACNA2D4}
  & 3 & 53 & 6.11E-02 & 12 & TRUE \\ hsa-mir-330 & \textit{MASP1} & 3 & 55 &
  1.96E-02 & 3 & TRUE \\ hsa-mir-190 & \textit{NPR2} & 3 & 4 & 7.38E-02 & 9 &
  FALSE \\ hsa-mir-361 & \textit{NTN4} & 3 & 42 & 2.20E-02 & 12 & TRUE \\
  hsa-mir-151 & \textit{PRKCE} & 3 & 384 & 5.83E-02 & 2 & TRUE \\
  \botrule
  \end{tabular*}}
  {}
  \end{table}

\subsection{Tool for interactive exploration of complete results}

We have presented only a few examples of gene-miRNA-SNP trios tested in
the cancer types shown above. In order to enable researchers to explore other trios,
we have produced an open source R Shiny application, called mirApp, that can be used to 
investigate other trios in the analyzed datasets. 
The user is asked to choose the cancer type (breast, liver, lung, or prostate)
and input a specific miRNA of interest.  The Shiny app then produces
a miRNA-specific manhattan plot of all significant regQTLs within
the cancer type ($p_{FDR}\leq0.1$). This manhattan plot is
interactive, such that when the user clicks on an individual regQTL 
point, the app produces its trio interaction
plot, similar to those in Figure~\ref{fig:BRCAtrios}. 
This tool can be downloaded freely from  
\url{https://github.com/gawilk/mirApp}.

In addition, code to carry out the analysis (to reproduce these results
or apply them to other SNP/miRNA/mRNA datasets) can be obtained from
\url{https://github.com/gawilk/miRNA-SNP}.

\section{Discussion}

We have described a novel integrative method that combines genomic 
and expression data to elucidate the effects that genomic variants exert on
miRNA regulation of genes in cancer. This integrative analysis combines 
miRNA expression, mRNA expression, and genotype data from tumor 
tissue to find polymorphisms that modulate co-expression patterns between 
miRNAs and their putative gene targets, which we term regQTLs. 
This analysis continues our previous work that identified miRNAs and 
entire pathways whose co-regulation was found to be disrupted in tumors. 
Here, we hone in on previously identified dysregulated pathways, and 
determine whether polymorphisms present within pathway genes may 
contribute to individual gene dysregulation. 

This work is in the spirit of other integrative omics analyses to yield insights 
into gene expression regulatory mechanisms. Its main novelty is to take into account
genomic variation and apply it on a genome-wide scale. 
Other integrative analyses of omics platforms have been applied to
yield discoveries on gene expression regulation mechanisms. Pipelines
such as CrossHub~\cite{krasnov2016crosshub} take into account miRNA and
TF relationships as well as methylation evidence, through TCGA and
ENCODE ChIP-Seq binding evidence, to describe regulation of gene
expression. RACER~\cite{li2014regression} uses regression analysis to
predict gene expression as a function of genetic and epigenetic
factors including copy number variation, miRNAs, DNA methylation, and
TF evidence combined from TCGA and ENCODE to study Acute Myeloid
Leukemia. Another study~\cite{setty2012inferring} combined similar
input variables in a linear fashion to model mRNA expression changes
in glioblastoma tumor samples, and was able to identify activities
that were predictive of subtypes and survival. These studies have
identified some relationships between expression regulators and genes
and focused on mostly single cancer types. Jacobsen and colleagues~\cite{jacobsen2013analysis}
used a statistical approach to model the recurrence of miRNA-mRNA
expression in tumor samples across multiple cancer types, induced by
changes in DNA copy number and promoter methylation. However, none of
these studies have taken into account genomic variation to address
their effect on gene regulation.

In contrast, our method incorporates genomic variation to 
identify regQTLs. Our method is fully data driven, integrating
sample specific expression and genomic data to find allele-specific regulatory
effects. By applying multiple linear regression models using all three omics features,
we can assess which SNPs differentially affect miRNA regulation of genes.
The use of linear models and ANOVA allow for relatively easy assessment 
of statistical interactions. Because we focus on genes within pathways 
found to be co-regulated with miRNAs which are disrupted in cancer, 
our approach may help find genomic variants that contribute to 
tumorigenesis. We emphasize that the application to dysregulated 
pathways permits the identification of regQTLs with potentially 
local and system effects, and significantly reduces the search
space of mechanisms under consideration in the genome.

We apply this analysis to breast, liver, lung, and prostate cancers, and within 
each cancer type, test millions of possible models (genes regulated by 
miRNAs modulated by SNPs, or ``trios''). We find polymorphisms 
systematically affecting miRNA-gene regulation, with many more 
statistically significant effects than expected by chance. This 
supports the notion that cancer contains significant perturbations
to the entire genome. Among the flagged trios with high significance,
many miRNAs and genes are often implicated in tumorigenic processes in the literature. 
These include tumor suppressor genes, genes in the the p53 network, genes within 
signaling pathways, and miRNAs whose aberrant expression or aberrant 
targeting has been documented in multiple cancer types. In addition, we find 
several genes each containing many individual variants differentially 
modulating miRNA regulation. These genes, and the genomic regions 
surrounding them, may indicate hotspots of tumorigenic 
interest for future research.

Due to the extensive nature of the study, an R Shiny app has been 
developed to fully explore all regQTLs and visualize their effects. Users 
can utilize the app interactively to observe regQTL significance 
genome-wide by cancer type, and plot individual miRNA-gene interactions
modulated by them. This utility allows for complete exploration of 
our integrative analysis of TCGA data.  

We note that the our results are somewhat limited by the input data. 
Currently, TCGA is the largest known resource of cancer omics data, 
with samples assayed across both expression and genotype data. 
However, individuals of European ancestry are highly overrepresented.
Having comparable datasets in diverse populations would strengthen the results 
of this study. In addition, our method only considers genes and SNPs which 
lie on annotated pathways; genes and SNPs that are currently 
unannotated on biological pathways, and therefore unconsidered in our model, 
may be of tumorigenic importance. Finally, our search for regQTLs is highly flexible; 
we do not restrict miRNA-gene relationships to those already corroborated 
with biophysical evidence, and, in addition, do not restrict genomic 
variants to those within putative miRNA binding regions. 
These criteria have been set to allow for novel discoveries, 
since computational miRNA-gene binding rules have been observed to deviate 
from experiment. However, we may also identify second-order effects 
of miRNA regulation modulated by genotype (or possibly spurious 
relationships) that would require functional experiments to elucidate.
Nevertheless, we do find significant relationships in 
which the genes and miRNAs are implicated in cancers in the literature. 

Our model is relatively simple and can be efficiently applied to
any combined miRNA/mRNA/SNP dataset of interest to reveal the
effects of a single regulatory SNP. 
We envision that future work could apply and extend our approach 
in several ways.  For instance, it is
conceivable that multiple SNPs in combination
will influence miRNA regulation of a gene, and that
genomic variation may affect other layers of
gene regulation (\eg, but influencing transcription factor binding).
Future extensions of this method could include 
integrating TF binding sites or epigenetic factors in the analysis regQTLs.
Given specific regQTLs identified in this study, other avenues could include 
validating their differential regulation by experimental means, or 
estimating their strength \textit{in silico}.  Perhaps the most exciting future application
would be to inform personalized medicine in the context of miRNA therapeutics~\cite{wang2009microrna,kasinski2015combinatorial,shah2016microrna}; for instance, the results shown in 
Figure~\ref{fig:BRCAtrios} suggest that targetting \textit{mir-190b} could influence
the expression of the tumor suppressor \textit{TUSC3}, but only amongst homozygous
AA individuals at regQTL rs13253051.

Finally, we note that our method for identifying regQTLs can be easily applied to other
diseases and experimental modalities (such as TFs) to determine the functional impact
of specific loci. A genome-wide analysis of functional regulatory effects can help identify 
polymorphisms and mutations that contribute to disease. 




\section{Acknowledgements}
The results published here are in whole or part based upon data generated by The Cancer Genome Atlas managed by the NCI and NHGRI. Information about TCGA can be found at \url{http://cancergenome.nih.gov}.

\section{Funding}
J.S. McDonnell Foundation (to RB); 
Northwestern University Data Science Initiative (to GW and RB).

\subsubsection{Conflict of interest statement.} None declared.

\clearpage
\bibliographystyle{unsrt} 
\bibliography{references}

\newpage
\section{Supplementary Information}

\subsection{Source code}

Source code for this analysis is available from \url{https://github.com/gawilk/miRNA-SNP} 

\subsection{SNP PCA plots}

We applied PCA to SNP genotype data to test for population substructure. Although TCGA 
heavily samples from European ancestry, there are other ancestry groups represented.
Nevertheless, most of the samples heavily cluster within the first two principal components, 
even in cancer types which have large sample sizes of other or unknown ancestry, suggesting 
little population substructure in the cohort. PCA was applied using the 
snpStats~\cite{snpStatsPackage} package in R. 

\begin{figure}[!h]
	\begin{minipage}{0.50\textwidth}
		\includegraphics[scale=0.42]{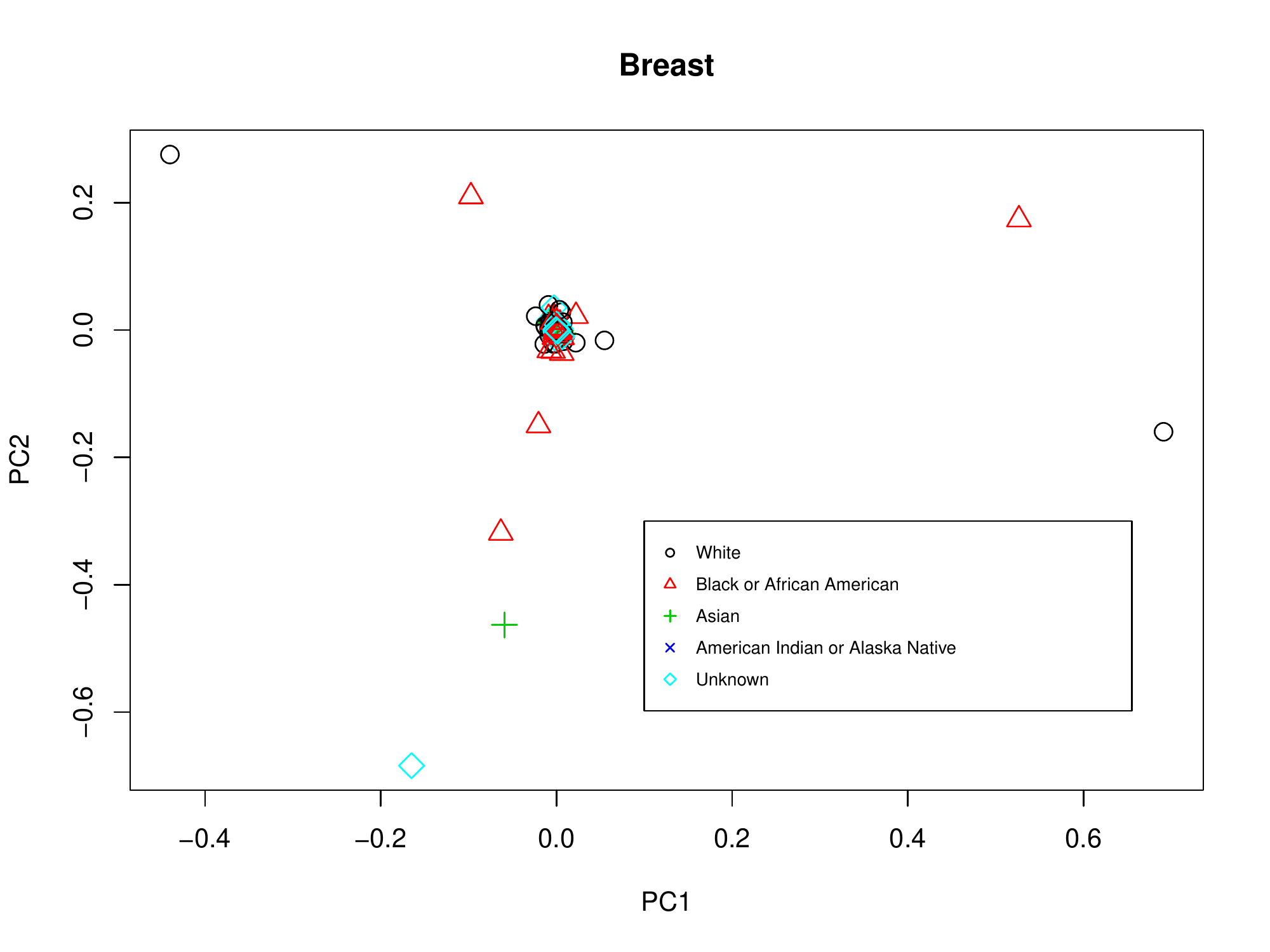}
	\end{minipage}
	\begin{minipage}{0.50\textwidth}
		\includegraphics[scale=0.42]{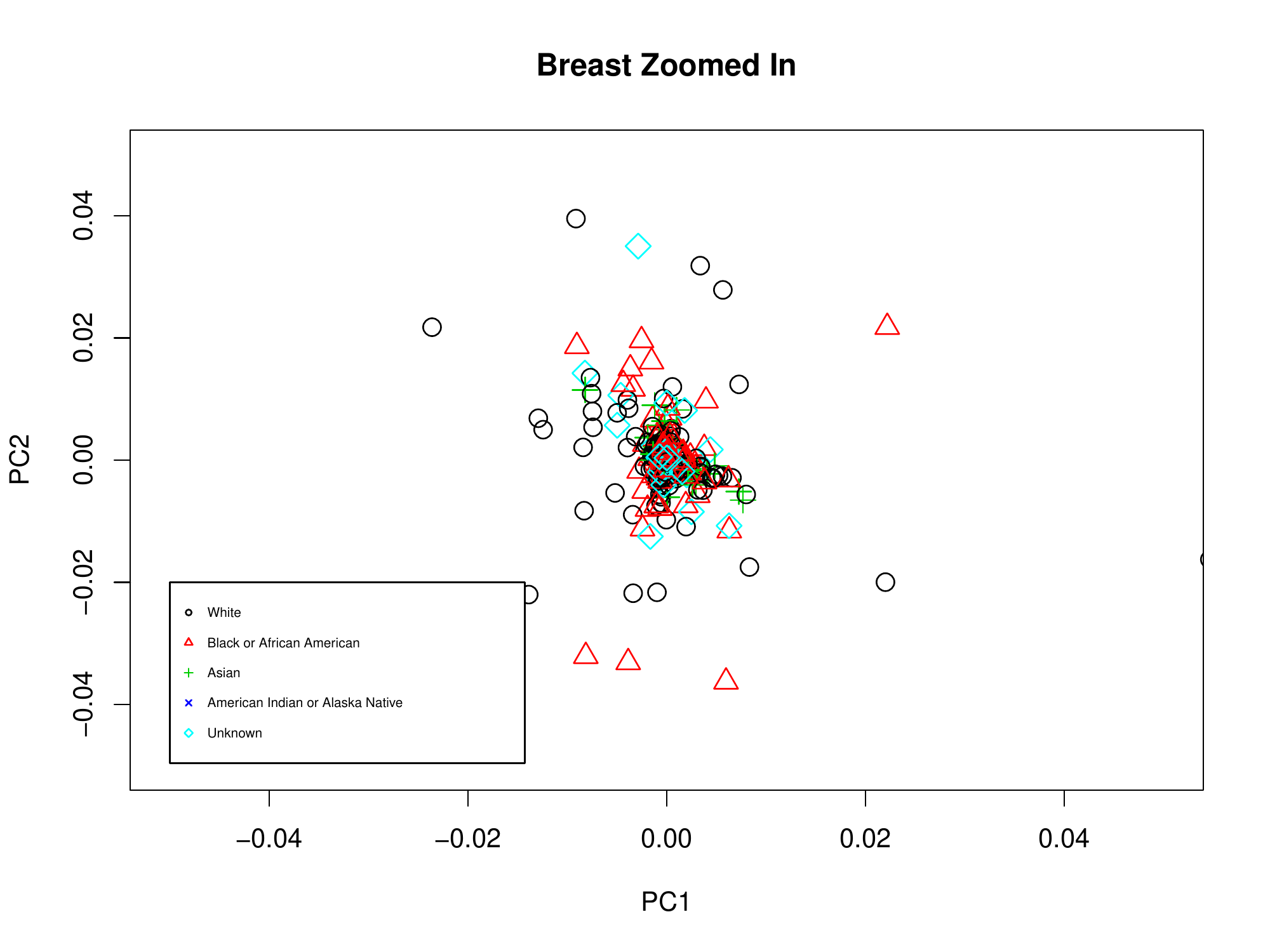}
	\end{minipage}
\caption{PCA plot of Breast SNP genotype data. The entire cohort (left plot) and the main cluster zoomed in (right plot) for the first two principal components. Sample groups appear to cluster together and exhibit little population substructure.} 
\label{fig:BRCA_PCA}
\end{figure}

\clearpage

\begin{figure}[!h]
	\begin{minipage}{0.50\textwidth}
		\includegraphics[scale=0.42]{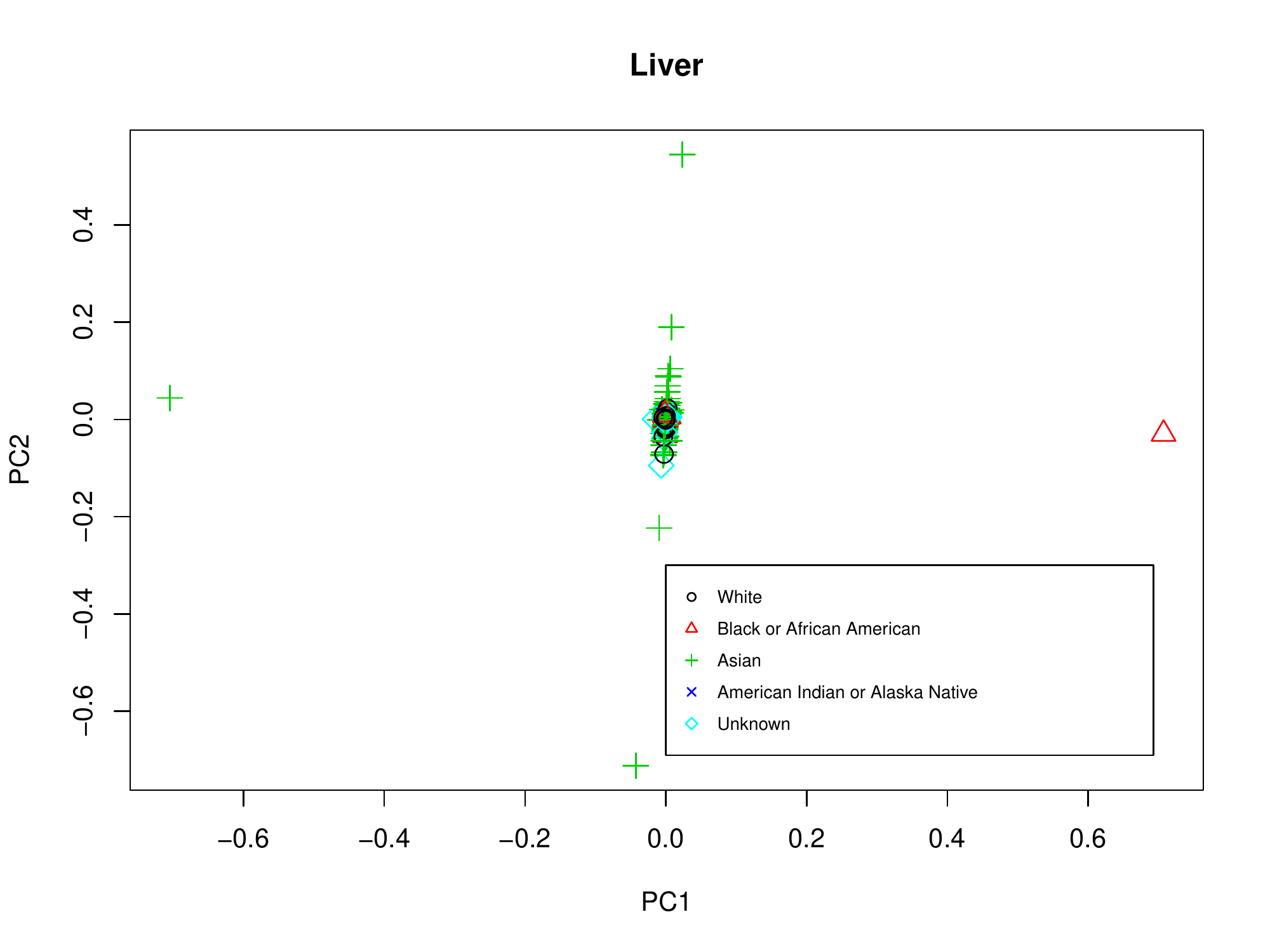}
	\end{minipage}
	\begin{minipage}{0.50\textwidth}
		\includegraphics[scale=0.42]{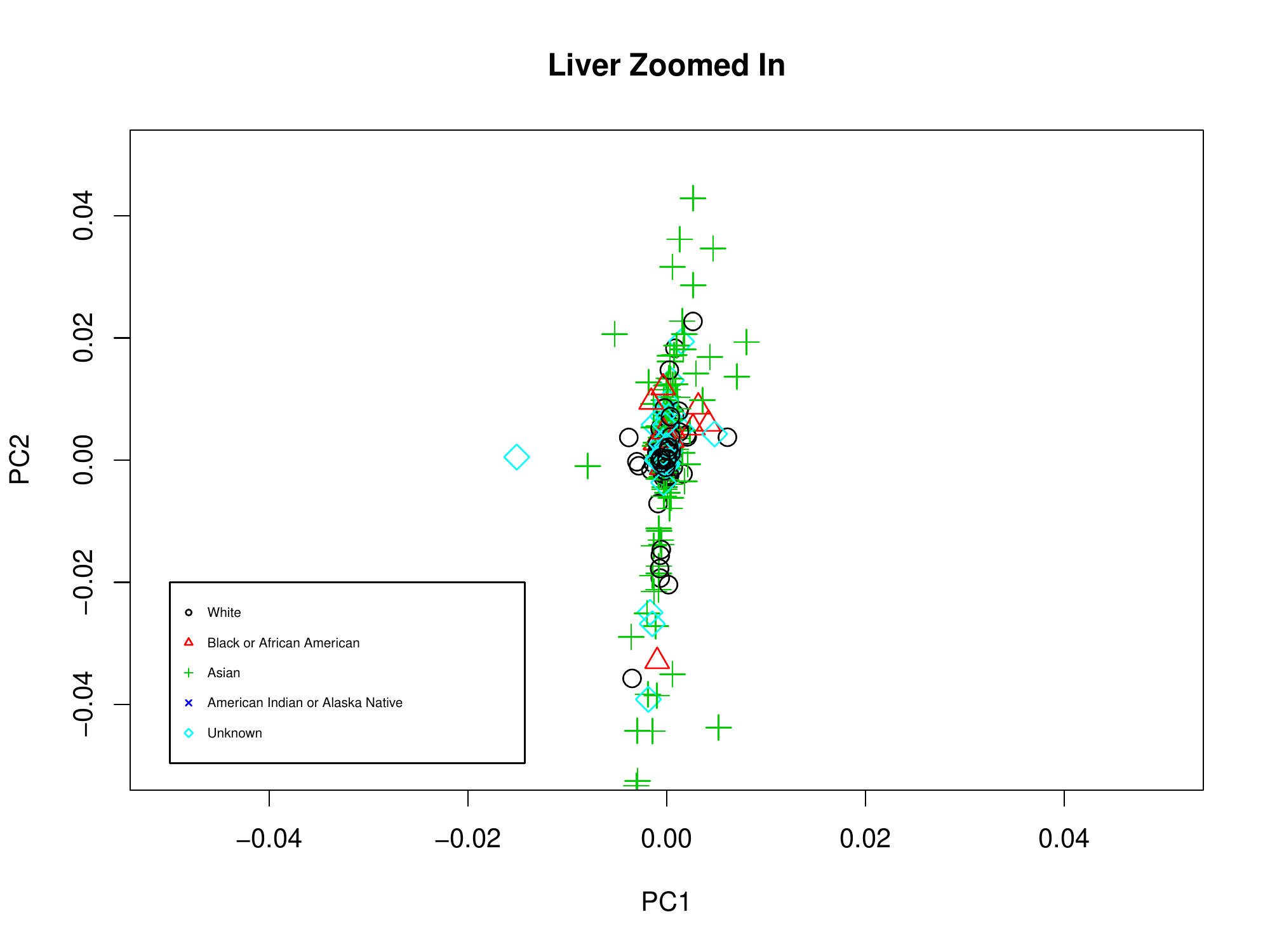}
	\end{minipage}
\caption{PCA plot of Liver SNP genotype data.} 
\label{fig:LIHC_PCA}
\end{figure}

\clearpage

\begin{figure}[!h]
	\begin{minipage}{0.50\textwidth}
		\includegraphics[scale=0.42]{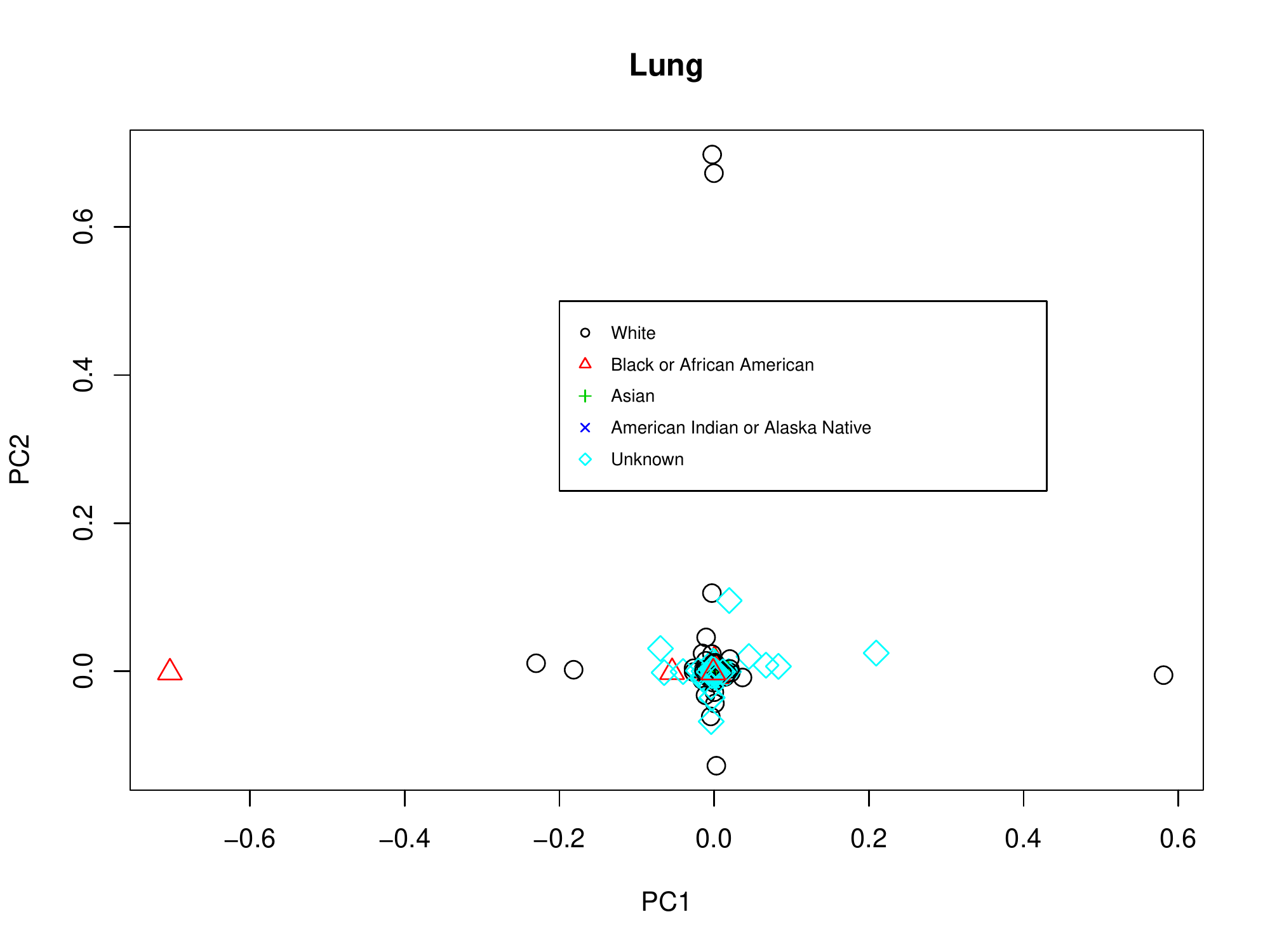}
	\end{minipage}
	\begin{minipage}{0.50\textwidth}
		\includegraphics[scale=0.42]{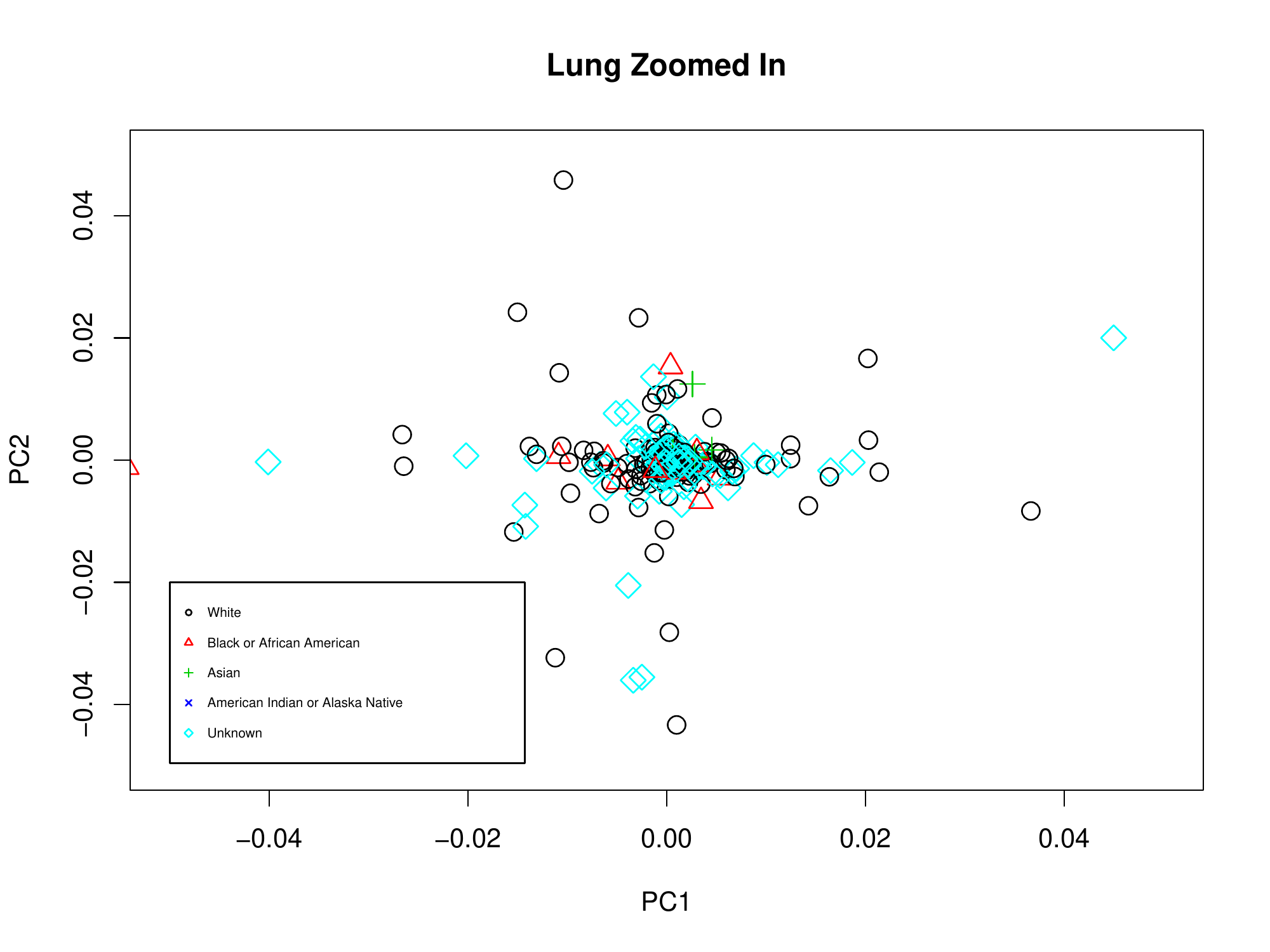}
	\end{minipage}
\caption{PCA plot of Lung SNP genotype data.} 
\label{fig:LUSC_PCA}
\end{figure}

\clearpage

\begin{figure}[!h]
	\begin{minipage}{0.50\textwidth}
		\includegraphics[scale=0.42]{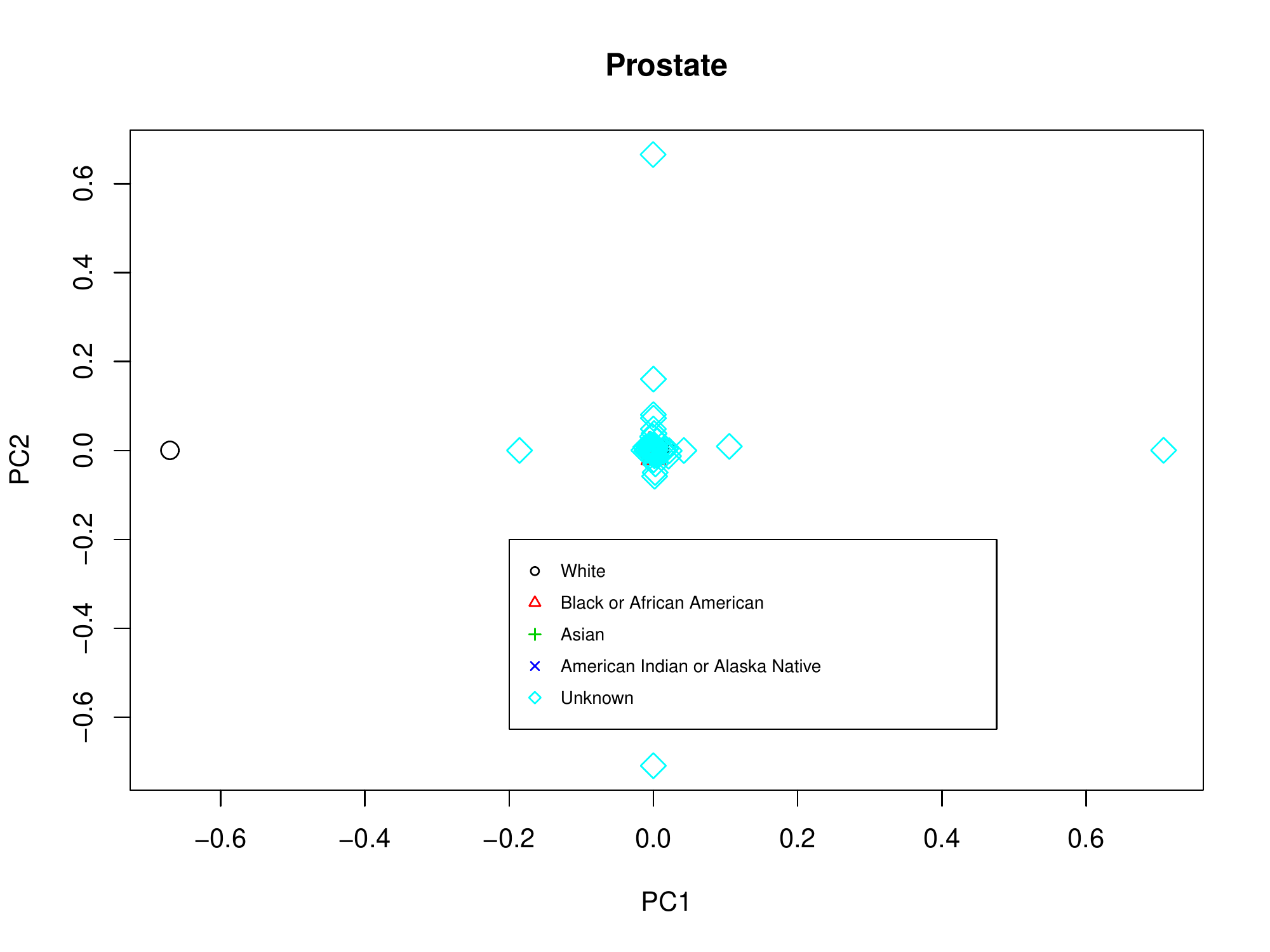}
	\end{minipage}
	\begin{minipage}{0.50\textwidth}
		\includegraphics[scale=0.42]{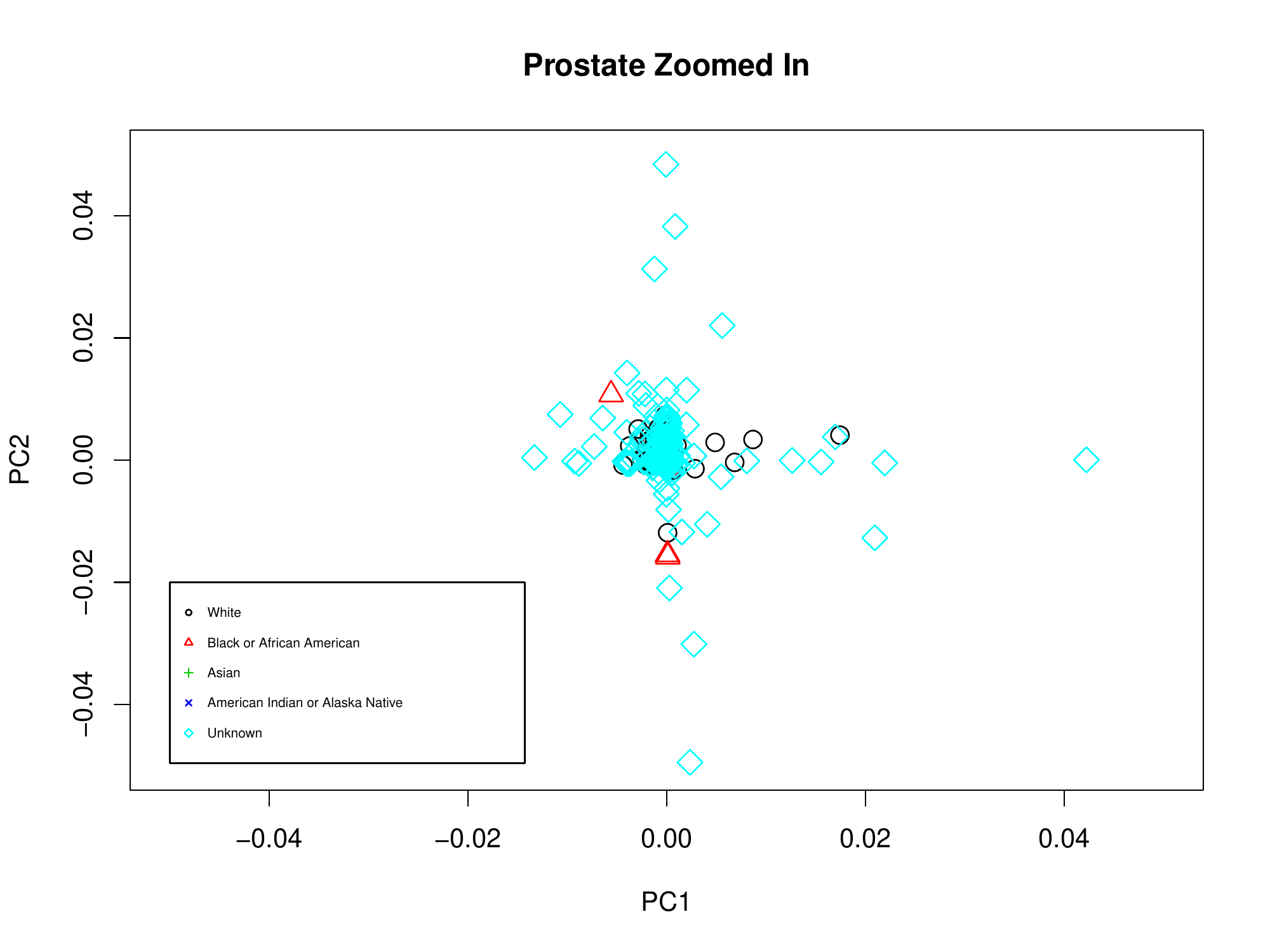}
	\end{minipage}
\caption{PCA plot of Prostate SNP genotype data.} 
\label{fig:LUSC_PCA}
\end{figure}

\end{document}